\documentclass[5p,times]{elsarticle}
\usepackage{url}

\usepackage{times}
\usepackage{dblfloatfix}
\usepackage{float}
\usepackage{caption}
\usepackage{multicol}
\usepackage{verbatim}
\usepackage{lipsum}
\usepackage{amsmath,amssymb,amsfonts,mathtools}
\usepackage{graphicx,wrapfig,picins}
\usepackage{textcomp}
\usepackage{xcolor,subfigure}
\usepackage{dblfloatfix}
\usepackage{float}
\usepackage{bbding}
\usepackage{lineno,hyperref}
\modulolinenumbers[5]

\journal{Journal of Future Generation Computer Systems}

\begin{document}

\begin{frontmatter}

\title{TAD: Transfer Learning-based Multi-Adversarial Detection \\ of Evasion Attacks against Network Intrusion Detection Systems}


\author[mymainaddress,mysecondaryaddress]{Islam Debicha\corref{mycorrespondingauthor}}
\cortext[mycorrespondingauthor]{Corresponding author}
\ead{islam.debicha@ulb.be}

\author[mymainaddress]{Richard Bauwens}
\author[mysecondaryaddress]{Thibault Debatty}
\author[mymainaddress]{Jean-Michel Dricot}
\author[mythirdaddress]{Tayeb Kenaza}
\author[mysecondaryaddress]{Wim Mees}

\address[mymainaddress]{ Cybersecurity Research Center, Université Libre de Bruxelles, 1050 Brussels, Belgium}
\address[mysecondaryaddress]{Cyber Defence Lab, Royal Military Academy, 1000 Brussels, Belgium}
\address[mythirdaddress]{Computer Security Laboratory, Ecole Militaire Polytechnique, Algiers, Algeria}

\begin{abstract}
	Nowadays, intrusion detection systems based on deep learning deliver state-of-the-art performance. However, recent research has shown that specially crafted perturbations, called adversarial examples, are capable of significantly reducing the performance of these intrusion detection systems. The objective of this paper is to design an efficient transfer learning-based adversarial detector and then to assess the effectiveness of using multiple strategically placed adversarial detectors compared to a single adversarial detector for intrusion detection systems. In our experiments, we implement existing state-of-the-art models for intrusion detection. We then attack those models with a set of chosen evasion attacks. In an attempt to detect those adversarial attacks, we design and implement multiple transfer learning-based adversarial detectors, each receiving a subset of the information passed through the IDS. By combining their respective decisions, we illustrate that combining multiple detectors can further improve the detectability of adversarial traffic compared to a single detector in the case of a parallel IDS design.
\end{abstract}


\end{frontmatter}

%\linenumbers

\section{Introduction}
With the growth of information communication and in particular the growth of the Internet, the concern to secure this information and related systems has become increasingly important. Every day, massive amounts of sensitive information are created, transferred, stored, or updated around the world. This sensitive information can range from personal emails to banking transactions, from a simple holiday photo to military communication. This information has always been the target of malicious entities wanting to steal, modify or delete it. To achieve these goals, hackers and other malicious agents have discovered, exploited, and improved a wide range of cyberattacks.

This new era of cyber security has required a paradigm shift from simple defense mechanisms to sophisticated defense systems. While basic network protections, such as firewalls, may have been sufficient in the past, the increasing complexity of cyberattacks has made them insufficient if used alone. To guard against these complex attacks, Intrusion Detection Systems (IDS) are now the cornerstone of cyber security.

As the sophistication of attacks increases, weaknesses are also discovered and exploited at an increasing rate. Attacks that exploit weaknesses almost as soon as they are discovered are called zero-day attacks. With zero-day hacking attacks breaking records year after year \cite{DBLP:journals/cybersec/Neillarchive}, it becomes necessary for IDSs to be able to detect never-before-seen attacks. Machine learning-based IDSs are one of the solutions to this problem. Indeed, numerous research papers have shown that the use of machine learning provides new approaches and detection mechanisms against zero-day attacks in all areas of cybersecurity \cite{DBLP:conf/cycon/ApruzzeseCFGM18, DBLP:journals/access/XinK0CLZGHW18, DBLP:journals/ijon/MahdavifarG19, DBLP:journals/access/VinayakumarASPA19}.

As always in the ongoing rivalry between cyber-attacks and defenses, as machine learning has become an important part of defense mechanisms, it has also become the target of attacks. The field of adversarial learning studies various methods of attack against machine learning and different ways of protecting models against these attacks. These new attacks take advantage of the fact that machine learning models have a discontinuous input-output mapping. This means that humanly imperceptible changes to the input of a model can cause a dramatic change in the classification result. These changes or perturbations in the input are called adversarial examples \cite{DBLP:journals/pieee/MillerXK20}.

This change in cyber-target has led to an equivalent change in cyber defenses, as the focus has shifted to protecting IDSs themselves from attacks created to undermine their effectiveness. While adversarial learning features some defenses, it is clear that more research is needed to refine these techniques. Adversarial detection has shown promising results in the field of computer vision, but very limited work has been done regarding this method in the field of intrusion detection systems. 

The main objective of this paper is to design and study the use of multiple strategically placed transfer learning-based detectors of adversarial attacks. The use of multiple detectors could lead to significantly better detection rates against adversarial attacks, as the information is distributed among them and the logical links between these pieces of information can be discovered independently. In our experimental work, the results show that the use of multiple adversarial detectors is more advantageous in parallel IDS than in serial IDS. This is mainly due to the fact that the detectors are more diversified in the parallel architecture. Our four main contributions are the following: (1) Implementation of two IDS models based on deep learning (one in serial form and the other in parallel form) and evaluation of the effect of four adversarial attacks on their performance. (2) Proposal of a new adversarial detection scheme based on transfer learning to allow a better information flow between the IDS and the adversarial detectors. (3) Assessment of the performance of the proposed approach when exposed to evasion attacks that were not seen in the training phase to simulate the zero-day attack scenario. (4) Performance evaluation of the use of multiple adversarial detectors versus a single detector in both serial and parallel IDS designs.

The remainder of the paper is organized as follows. Preliminaries and related works are presented in Section \ref{sec:Pre}. We present our evaluation methodology in Section \ref{sec:det}. We present our performance evaluation in Section \ref{sec:fus}. We conclude in Section \ref{sec:cfw}.

\section{Preliminaries \& Related Works}
\label{sec:Pre}

%As intrusion detection has become a more relevant defense against network attacks, different types of intrusion detection systems have been developed, each with its advantages and disadvantages. From the type of information processed to the way in which that information is processed, the different types of IDS vary considerably, and an appropriate taxonomy is needed. Like intrusion detection, adversarial learning is a growing field that has developed in recent years. Attacks may vary in their objectives or in the methods used to achieve those objectives, resulting in a wide range of defense responses.

This section first presents a brief description of the concepts of adversarial learning, as well as the metrics used to evaluate the performance of adversarial attacks. We then present the attacks used in this paper and their methodology. Similarly, we describe the defense mechanisms that can be implemented to impede adversarial attacks. This section also presents the transfer learning techniques used in our detection mechanism. Furthermore, when considering multiple detectors, it is necessary to find a way to combine their respective evaluations into a final decision. Thus, this section presents the three fusion rules used in this work.

\subsection{Adversarial Learning}

Adversarial learning is the name given to the problem of devising attacks against machine learning as well as the defenses against those attacks \cite{DBLP:journals/pieee/MillerXK20}. There exists a large number of attacks against machine learning models that are often classified with regard to their respective goals \cite{DBLP:journals/isci/CoronaGR13}. Depending on the phase in which the attack is conducted, adversarial attacks can be divided into either poisoning or evasion attacks.

One of the main differences between poisoning and evasion attacks is the timing of the attack. In the case of the poisoning attack, the adversary targets the IDS during the training phase, whereas for the evasion attack, the adversary launches its attack on the IDS during the testing phase (i.e., only after the IDS has been trained and deployed). Clearly, the poisoning attack is more difficult to execute in a real-world setting, as it requires the attacker to re-train the IDS with its poisoned data. Evasion attacks, on the other hand, allow the attack to be performed without modifying the IDS. 

Since evasion attacks are more practical and therefore more widely used against IDSs, our work will focus on evasion attacks and defenses against them. In particular, this work will only consider evasion attacks against Deep Neural Network-based IDS (DNN-IDS). Although DNNs seem to be the most promising research area in terms of IDS performance \cite{DBLP:journals/access/LanskyAMMKRHR21}, these models also seem to be the most vulnerable to adversarial attacks \cite{DBLP:journals/pieee/MillerXK20}.

\subsubsection{Assumption of adversary knowledge}

When attacking a system in a controlled environment, the first thing to decide is the adversary's knowledge. Indeed, we distinguish between an attacker with the full knowledge of the to-be-attacked model as well as any defenses in place (white-box) and an attacker with no prior knowledge of the attacked system (black-box).

In the case of IDSs, knowledge of the initial classifier is plausible because most IDSs use state-of-the-art models and learning methods. Knowledge of the defenses, on the other hand, is less plausible because there is no clear consensus on the best defense mechanism to implement in this case. In this work, we assume a kind of gray-box situation where the attacker has full knowledge of the IDS but no prior knowledge of the defenses that are employed to defeat it.

\subsubsection{Metrics}

When performing adversarial attacks, several metrics can be considered to assess the results. Indeed, one of the commonly used metrics is the success rate of the attack: the difference between the performance of the model before and after the attack. The other commonly considered metric is the strength of the attack. Some attacks allow a strength metric to be defined, which increases the success rate but also potentially the detectability of the attack. The strength often refers to the maximum perturbation that can be applied to the initial sample. 
In this paper, we evaluate the model performance using the standard machine learning metrics: Precision, Recall, and F1-score to evaluate the performance of the IDSs and Detection Rate (Recall) to assess the performance of the adversarial detectors. These metrics are calculated as follows:
\begin{equation}
\text{	Precision} = \frac{tp}{tp + fp} 
\end{equation}
\begin{equation}
	\text{Recall} = \frac{tp}{tp + fn} 
\end{equation}
\begin{equation}
	\text{F1-score} = 2 \cdot \frac{Precision * Recall}{ Precision + Recall}
\end{equation}

\subsection{Adversarial attacks}
\label{attacks}
%When a DNN is trained to perform a certain task, the objective is to gradually approach a correct solution, represented by the label in supervised learning. To quantify the distance between the correct solution and the one given by the DNN during training, a mathematical function is used. This function is called the loss function, and the whole purpose of learning is to minimize this function, thus getting closer to the correct solution.
%
%Each learning step consists of adjusting the internal weights and biases of the DNN according to the feedback given by the value of the loss function. To adjust these values, learning is done in a rolling ball manner, by making small adjustments in the direction that brings the loss function closer to a local minimum. To find this direction, the negative gradient is used. Indeed, the gradient is the direction of the steepest ascent, using its opposite gives the direction of the steepest descent hence getting closer to a minimum. This process of small consecutive adjustments along the negative gradient is called back-propagation.

\subsubsection{Fast Gradient Sign Method (FGSM) \cite{DBLP:journals/corr/GoodfellowSS14}}

In an FGSM attack, the objective is to maximize the classification error by maximizing the loss function. Since the attacker cannot modify the internal weights and biases of the attacked model, he instead modifies the input. These modifications are applied with the goal of maximizing the loss function of the victim model. In order to maximize this loss function, the alteration made to the original input vector is calculated with respect to the gradient (i.e., the direction of the steepest ascent) of the original model. To be able to calculate this gradient, FGSM requires internal knowledge of the attacked model, namely the weights and biases. A specially crafted perturbation is applied to each feature following the sign of the gradient.

%In mathematical terms, this means that the modified adversarial example $Adv_x$ would be given by Eq. \ref{eq:FGSM} : 
%\begin{equation}\label{eq:FGSM}
%	adv_x = x + \epsilon * sign(\nabla_x J(\theta, x, y))	
%\end{equation}
%
%Where $x$ is the initial input, $J$ is the loss function, $y$ is the original input's label, and $\theta$ is the model parameters. $\epsilon$ is a small multiplier, to ensure that the perturbations remain small.

%While most research on FGSM is performed in a white-box scenario, which means that the adversary has complete knowledge of the classifier and its internal structure, in \cite{DBLP:conf/milcom/YangLZF18} the authors investigate the possibility of performing such an attack on a black-box model. They obtained significant results, showing that performing such an attack without any internal knowledge of the classifier is possible under some conditions.

\subsubsection{ Projected Gradient Descent (PGD) \cite{DBLP:conf/iclr/MadryMSTV18}}

PGD is an iterative variant of the FGSM attack. The method uses iterative applications of the FGSM method, with a smaller step size. This iterative idea ensures that the resulting adversarial samples are minimally modified while being theoretically misclassified by the attacked classifier. The use of the iterative method is computationally expensive compared to FGSM, so a trade-off must be made between efficiency and computational power.

%Mathematically, this iterative method can be described by Eq. \ref{eq:PGD} :
%\begin{equation}\label{eq:PGD}
%	x^{t+1} = \Pi_{X+\epsilon}(x^{t} + \alpha *sign(\nabla_x J(\theta, x, y)))
%\end{equation}
%
%
%Where $\alpha$ is the step size, meaning the maximum alteration that can be applied to a feature at each step of the attack. The $\Pi$ function is a per-feature clipping function that ensures that each feature will remain in the $\epsilon$-ball from its original value. Both $\epsilon$ and $\alpha$ are hyperparameters that control the amount of perturbation.

\subsubsection{DeepFool (DF) \cite{DBLP:conf/cvpr/Moosavi-Dezfooli16}}

In the paper presenting the DF attack, the authors introduce the notion of robustness for binary classifiers at a given point. This robustness would be quantified by the minimal perturbation introduced to this point in order to change its classification. This perturbation is equal to the distance between the original point and the closest decision boundary of the classifier. Once the robustness is estimated, the final perturbation vector $r$ is multiplied by a constant of the form $1 + \epsilon$ in order to cross the decision boundary. This $\epsilon$ can be considered as a meta-parameter of attack strength.

%In the case of an affine binary classifier, $f(x) = w^T x + b$, this decision boundary can be described as the affine hyperplane $F = \{x : w^T x + b = 0\}$. The minimal perturbation to change the classification of a certain $x_0$ would then be the orthogonal projection of $x_0$ onto $F$. This perturbation, $r_*$, is given by the closed-form formula, as shown in Eq. \ref{eq:DF}:
%\begin{equation}\label{eq:DF}
%	r_*(x_0) = - \frac{f(x_0)}{||w||_2^2}w
%\end{equation}
%
%
%In the case of non-affine binary differentiable classifiers, the attack adopts an iterative method. At each step, the classification function $f$ is linearized around the point $x_i$ and the minimal perturbation of the now linearized classifier is computed using the same formula as for an affine classifier. By applying this perturbation to $x_i$, we obtain the next point $x_{i+1}$ and we can perform the linearization step again. The attack stops when $x_{i+1}$ has a different classification than $x_0$. The robustness (and thus minimal perturbation to be brought to $x_0$) can then be estimated by summing all the individual perturbations $r_i$.

\subsubsection{Carlini \& Wagner (CW) \cite{DBLP:conf/sp/Carlini017}}

While having the same goal as the FGSM attack, the CW attack uses a different approach. Here, the optimization problem is formulated in a new way. That new way of formulating the problem makes it solvable by using an optimization algorithm.
% The formal definition of the optimization problem is as shown in Eq. \ref{eq:CW}:
%\begin{equation}\label{eq:CW}
%	\begin{multlined}
%		\quad	minimize :
%		\quad D(x, x+\delta) \\
%		\quad such \ that :
%		\quad C(x+\delta) = t
%		\quad x+\delta \in S
%	\end{multlined}
%\end{equation}

This implies finding a perturbation $\delta$ that minimizes the distance metric $D$ while changing the classification of $x+\delta$ from the original classification of $x$ to the desired class $t$. This $\delta$ alteration must also ensure that the adversarial example still is in the set $S$ of accepted samples. This clause limits the perturbation of the features so that it does not render the initial network attack invalid.

%To make this optimization problem solvable, the authors redefine an objective function $f$ so that $C(x + \delta) = t$ iff $f (x + \delta) \leq 0$. While the authors propose several objective functions, in \cite{DBLP:journals/fgcs/PawlickiCK20} the $L_2$-norm version was used. That version uses an objective function defined as shown in Eq. \ref{eq:CW2}
%\begin{equation}\label{eq:CW2}
%	f(x') = max(max\{Z(x')_i : i \neq t\} - Z (x')_t, -k)	
%\end{equation}
%
%
%In this equation, the parameter $k$ allows the attacker to control the confidence with which the sample will be classified as class $t$.

\subsection{Defenses}

According to Miller et al.\cite{DBLP:journals/pieee/MillerXK20}, defenses against evasion attacks are divided into two categories: robust classification and anomaly detection. 
\subsubsection{Robust classification}
The goal of the robust classification strategy is to correctly classify any adversarial examples, without compromising test performance in the absence of adversarial attacks. To achieve this, the training (and/or testing) process is typically modified (e.g., by preprocessing the data, using a robust training objective, or training with adversarial examples).

One of the earliest defense mechanisms proposed in the literature to counter evasive attacks is adversarial training \cite{DBLP:journals/corr/SzegedyZSBEGF13,DBLP:journals/corr/GoodfellowSS14, DBLP:journals/tnsm/ApruzzeseAMVC20}. Adversarial training is a method of data augmentation, which involves adding data to the training set used to build the classifier. This added data consists of adversarial examples created against the model. These adversarial samples are assigned the correct label so as to create, in theory, a more robust classifier that learns to correctly classify samples even if they have been modified by an adversarial attack.

%Apruzzese et al.\cite{DBLP:journals/tnsm/ApruzzeseAMVC20} proposed to strengthen flow-based botnet detectors through adversarial training by exploiting deep reinforcement learning to generate realistic adversarial samples and then using these samples to strengthen the detector through adversarial training. The authors showed that their defensive scheme increased the detection rate of the studied attacks and that the performance of their botnet detectors did not degrade in non-adversarial settings.

%Most of the time, adversarial training requires a large number of adversarial examples to train the classifier on all possible sample alterations and attack variations. Therefore, in most real-world cases, a worst-case perturbation is estimated \cite{DBLP:journals/pieee/MillerXK20}. Given this worst-case perturbation, the adversarial examples will be designed to train the model to be robust against it. This stems from a common assumption that if the model is robust against the worst-case scenario, it is likely to be robust against the more common cases as well.

Although this method gives some promising results in certain research areas, it is vulnerable to the blind-spot problem \cite{DBLP:journals/pieee/MillerXK20}. Indeed, if adversarial examples of a certain attack family were not added to the training set, the classifier might not be robust against them. It also means that this defense mechanism may not be as effective against zero-day attacks, which are one of the main concerns of the IDS field. Furthermore, as the IDS is retrained with adversarial examples, this could lead to a degradation of its performance in its initial classification task, as shown in \cite{DBLP:journals/corr/abs-2104-09852}.

Another adversarial defense proposed in the literature is defensive distillation\cite{DBLP:conf/sp/PapernotM0JS16, DBLP:journals/tetci/ApruzzeseACM20, DBLP:conf/esorics/GrossePMBM17} which was initially employed to reduce the dimensionality of DNNs, a distillation-based defense is proposed by Papernot et al.\cite{DBLP:conf/sp/PapernotM0JS16} where the objective is to smooth the decision surface of the model. In \cite{DBLP:journals/tetci/ApruzzeseACM20}, Apruzzese et al. introduced a defensive scheme to limit the impact of adversarial disturbances by employing the defensive distillation technique on random forest-based IDS. Experimental evaluations demonstrated the effectiveness of the proposed method with respect to the literature. The authors also found that their defensive scheme does not affect the performance of the IDS in non-adversarial settings.
Carlini and Wagner \cite{DBLP:journals/corr/CarliniW16} showed that defensive distillation is not secure, i.e., it is no more resistant to targeted misclassification attacks than unprotected neural networks. It is therefore unclear whether the adapted version of defensive distillation for the random forest is more secure against strong attacks.

An alternative approach to countering adversarial attacks is to negate the impact of changes applied to a given feature by ruling it out. Thus, as a defense against a broad selection of adversarial attacks, one could disregard all features that are susceptible to being manipulated by the attacker \cite{DBLP:journals/tcyb/ZhangCBYR16, DBLP:conf/acsac/SmutzS12} without changing the logic of the network traffic. However, Apruzzese et al \cite{DBLP:conf/nca/ApruzzeseCM19} showed that doing so has a detrimental impact on the performance of detectors in non-adversarial settings, e.g., by triggering more false alarms, which is highly undesirable for modern cyber security platforms. The reason for this reduction in quality is that the removed features have a significant impact on the underlying mechanisms of the baseline detectors; as such, excluding them results in a significant drop in performance, with most detectors averaging well below acceptable scores in real-world settings.

\subsubsection{ Anomaly	detection}
\label{sec:AnomalyDetection}
A considerable amount of research has been done in recent years on an alternative defense strategy against evasion attacks which is anomaly detection. One of the reasons given is the fact that robust classification of attacked samples is challenging, whereas their detection is somewhat easier \cite{DBLP:conf/ccs/Carlini017}. One supporting argument is that if a good robust classifier is conceived, then a good detector is obtained for free - detection can be performed when the decision of the robust classifier disagrees with the decision of the non-robust classifier \cite{DBLP:journals/pieee/MillerXK20}. Our work falls into the category of anomaly detection.

This method proposes that instead of trying to classify the adversarial instance correctly despite the attack, the adversarial instance is detected and dealt with accordingly. Most often, this action consists in rejecting the detected sample and not letting it pass through the system\cite{DBLP:conf/iccv/LuIF17,DBLP:journals/neco/MillerWK19, DBLP:journals/corr/abs-2112-12095} 

Many detection methods in the area of adversarial learning have been proposed. In \cite{DBLP:journals/neco/MillerWK19}, the authors used the neural activation pattern to detect adversarial perturbations in the image classification domain. A method similar to theirs was also used in \cite{DBLP:journals/fgcs/PawlickiCK20} to detect adversarial traffic in the IDS domain. The method consists of taking a set of test samples (consisting of adversarial examples and clean samples) and getting the to-be-defended model to classify them. During this classification process, a neural activation vector is extracted for each sample. With these vectors, along with the initial sample label (adversarial or clean), a new labeled dataset is created. Using this new dataset, a detection model is trained to learn which activation patterns are most likely to be observed when classifying an adversarial sample.

It is worth mentioning that other works have studied the use of ensemble techniques to improve the detection rate against adversarial attacks, such as  \cite{DBLP:journals/tetci/ApruzzeseACM20}, \cite{DBLP:journals/tnsm/ApruzzeseAMVC20} and \cite{DBLP:journals/symmetry/ApruzzeseAMCR20}, although in these papers the use of ensemble techniques has a different meaning than the one we consider in this paper. For them, the ensemble technique is to train each classifier on a specific attack or application to ensure proper tuning for that task, but this might lead to less generalization in the training process, making it difficult to handle unknown attacks. The overall scheme for them is to assemble multiple classifiers, each specialized in a particular attack or application, whereas for ours, each instance is inspected by all components of the ensemble scheme leading to a deeper inspection of each particular instance. It should also be noted that approaches \cite{DBLP:journals/tetci/ApruzzeseACM20}, \cite{DBLP:journals/tnsm/ApruzzeseAMVC20} and \cite{DBLP:journals/symmetry/ApruzzeseAMCR20} seek to make the classifier robust by adding and/or filtering the training data used to train the model. This is not always possible, especially in the case of an already trained and deployed IDS. Our approach, on the other hand, does not affect the IDS but rather inspects it using other subnets, allowing this defense to be applied to already trained and deployed IDSs.

A similar design to that suggested in this paper can be found in \cite{DBLP:conf/iclr/MetzenGFB17} where the authors present an approach to detect adversarial perturbations in images by attaching a small sub-network to a deep neural network. This sub-network (detector) is trained specifically to detect adversarial perturbations. According to the authors, the detector performs surprisingly well on CIFAIR10 \cite{krizhevsky2009learning} and ImageNET \cite{DBLP:conf/cvpr/DengDSLL009} datasets (both are image classification datasets). Their approach is to train a single detector using  Feature Extraction as a transfer learning technique. The authors did not provide a way to determine the optimal positioning of the detector, rather they tried different positions where the single detector could be attached. In contrast, our work consists of designing an adversarial detector based on duplication + Fine-Tuning as a transfer learning technique. We show that this gives better results than Feature Extraction (As we will see in Section \ref{sec:compADS}). To further improve the detection rate, we propose and investigate an alternative to a single detector by using multiple strategically placed detectors combined with a sophisticated fusion rule.

\subsection{Transfer Learning}

Many real-world deep learning configurations involve the need to learn new tasks without compromising the performance of existing tasks. For instance, an object recognition robot may be shipped with a default range of abilities, but additional object models that are specific to the given environment need to be considered. In order to achieve this, new tasks should ideally be learned by sharing the parameters of old tasks without degrading the performance of the latter \cite{DBLP:journals/pami/LiH18a}.

\begin{figure*}[bh!]
	\centering
	\includegraphics[width=0.8\linewidth]{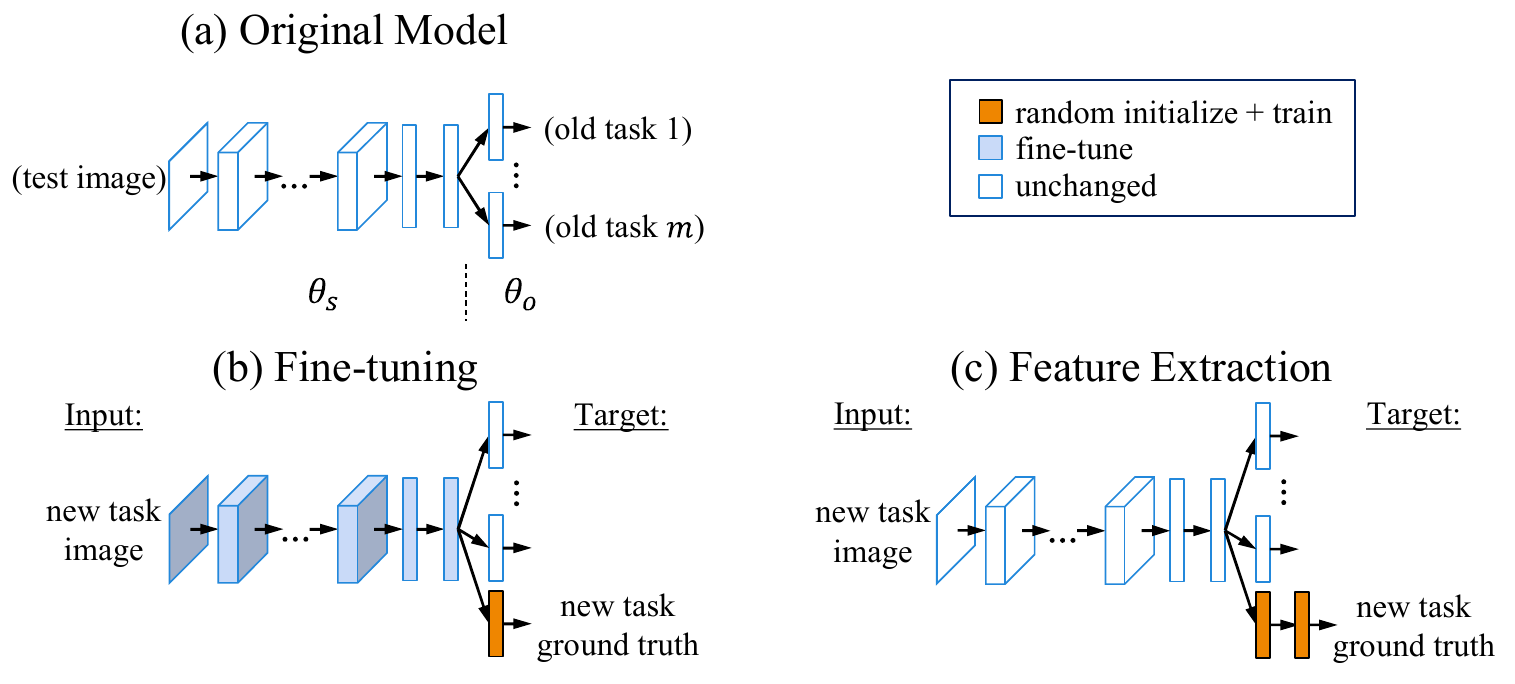}
	\caption{An illustration of two transfer learning techniques:Fine-Tuning and Feature Extraction, from \cite{DBLP:journals/pami/LiH18a}}
	\label{fig:ftfe}
\end{figure*}

Given $\theta_{s}$ a set of shared parameters for a DNN and $\theta_{o}$ task-specific parameters for previously learned tasks, three typical approaches exist for learning new task-specific parameters, $\theta_{n}$, while leveraging previously learned $\theta_{s}$ as shown in Fig. \ref{fig:ftfe}:

\begin{itemize}
	\item Feature Extraction (FE): when learning $\theta_{n}$, we keep $\theta_{s}$ and $\theta_{o}$ unaltered, and the outputs of the shared layers are used as features for the new task learning.
	
	\item Fine-Tuning (FT): in this case, $\theta_{s}$ and $\theta_{n}$ are tuned for the new task, whereas $\theta_{o}$ is unchanged. 
	
	\item Duplication + Fine-Tuning (DFT): it is possible to mirror the original deep neural network and fine-tune it for each new task to build a dedicated neural network.
	
\end{itemize}

\subsection{Fusion Rules}
\label{FR}

In order to combine the individual decisions of each detector involved in our proposed design, we will use three fusion rules: majority voting, Simple Bayes averaging, and Dempster-Shafer combination \cite{DBLP:journals/access/MohandesDA18}.

The majority vote rule is a simple fusion rule and will serve primarily as a reference for the other two rules in our work. With this rule, the final decision will simply be the individual decision made by the majority of the detectors. The main limitation of this fusion rule is that it processes all models equally, meaning that all models contribute equally to the prediction. This poses a problem if certain models perform well in some situations and poorly in others.

When performing a Bayesian average, the activation score for each class will be summed over all the detectors and then divided by the number of detectors. The highest resulting average will be chosen as the final classification by the system. Using this method helps to account for the confidence of each detector in its classification.

The final fusion rule used in this work is the Dempster-Shafer rule of combination \cite{Dempster}. This rule combines evidence elements from different sources (e.g., detectors) to achieve a degree of belief for each hypothesis (classification decision in our case). 

let $\Omega=\{\omega_1,...,\omega_K\}$, and $\mathcal{P}(\Omega)=\{A_1,...,A_Q\}$ is its power set, where $Q=2^K$. A mass function $M$:$\mathcal{P}(\Omega)$ $\rightarrow$ $[0,1]$ is a basic belief assignment (\textit{bba}) if $M(\emptyset)=0$ and $\sum_{A \in \mathcal{P}(\Omega)}{M(A)}=1$. 

In case where two \textit{bba}s $M_1$ and $M_2$ denote two elements of evidence (e.g., information from two detectors), we can combine them together using the Dempster-Shafer fusion rule, which results in $M=M_1\oplus M_2$ that is defined by Eq. \ref{eq:DSF}.
\begin{equation}\label{eq:DSF}
	M(A)=(M_1 \oplus M_2)(A)\propto \sum_{B_1\cap B_2=A}{M_1(B_1)M_2(B_2)}
\end{equation}

%An evidential \textit{bba} $M$ can be converted into a probabilistic one using Smets’ technique \cite{DBLP:journals/ai/SmetsK94}, where every belief mass $M(A)$ is evenly distributed over all elements of $A$, resulting in the so-called ``pignistic probability'', $Bet$ , given by Eq. \ref{eq:pignistic}.
%
%\begin{equation}\label{eq:pignistic}
%	Bet(\omega_i)=\sum_{\omega_i\in A\subseteq \Omega}{\frac{M(A)}{|A|}}
%\end{equation}
%where $|A|$ is the number of elements of $\Omega$ in $A$.

\section{Evaluation methodology}
\label{sec:det}
Anomaly detection is one of the key defenses proposed against adversarial learning in intrusion detection systems. Indeed, to perform any variant of evasion attacks on a trained model, the original features used by the model are modified to maximize the classification error. These modifications can be detected in a variety of ways, including using another neural network trained specifically to recognize the modified input packets. These new neural networks are called adversarial detectors (AdD).

%Just as there exist multiple ways to implement intrusion detection systems, there exist multiple ways to implement adversarial detectors and train them. Most of the time, a single adversarial detector is placed at a key step of the intrusion detection system and trained on the entirety of the feature set passed to the detection system.

This work aims to investigate the effectiveness of multiple adversarial detectors working together to detect adversarial perturbations during the inference phase. Each of these detectors is placed under a sub-network to receive different levels of information from the DNN-IDS. By combining their respective decisions, we could obtain a final classification affirming the legitimacy of each sample. The hypothesis here is that due to the added level of granularity, multiple detectors could help detect a wider range of perturbations and thus a wider range of evasion attacks. In addition, the use of transfer learning techniques would allow adversarial detectors to learn some important patterns from the IDS (since the two tasks are similar) while augmenting their knowledge based on the adversarial patterns.

%Since we are playing the role of the defenders, our proposal needs to resist strong attacks. Thus, the assumptions made regarding the capabilities of the attacker are those of a white-box attack: the attacker has full knowledge of the DNN-IDS (e.g., its gradient) and can modify the features passed to the DNN-IDS directly. We also put no limit on the computational and time resources of the attackers.

\begin{figure}[h!]
	\centering
	\includegraphics[width=1\linewidth]{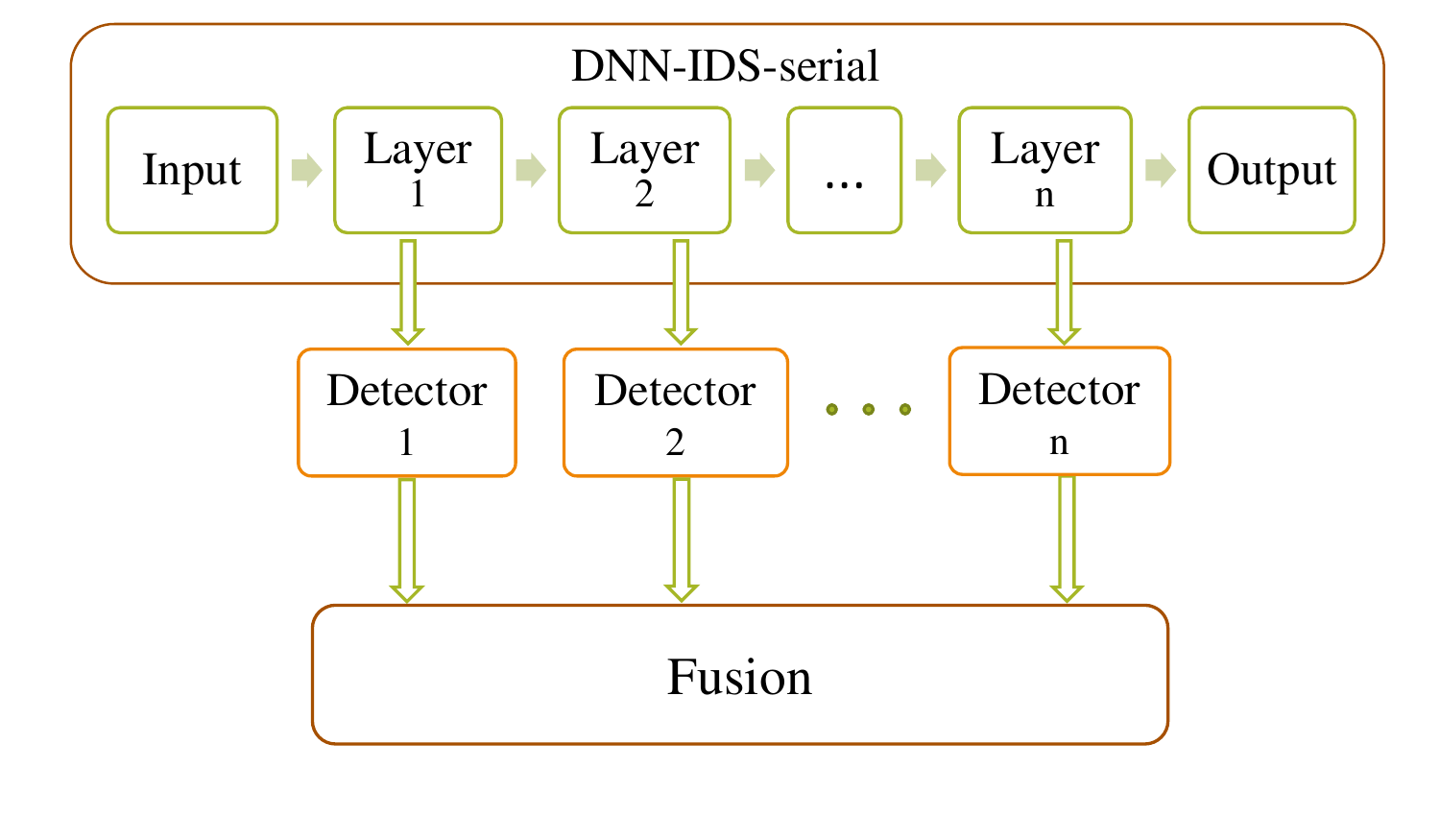}
	\caption{An illustration of the adversarial detectors in a serial DNN-IDS design}
	\label{fig:schema1}
\end{figure}

To do this, we build two DNN-IDS in both serial (DNN-IDS-serial) and parallel (DNN-IDS-parallel) architectures, perform multiple attacks against these two IDSs, and then design our adversarial detectors by following several transfer learning techniques to finally combine their decisions as shown in Figures \ref{fig:schema1} and \ref{Fig:minato}. This defense technique is then compared to adversarial learning.

\subsection{Network traffic datasets}
The experimental evaluation considered in our paper is performed on two public network traffic datasets, NSL-KDD\cite{DBLP:conf/cisda/TavallaeeBLG09} and CIC-IDS2017\cite{DBLP:conf/icissp/SharafaldinLG18}. NSL-KDD is chosen in this work because of its frequent use in the literature as a benchmark to compare different intrusion detection methods. CIC-IDS2017, on the other hand, is chosen because of its recency, which ensures that it can be considered a good representation of large modern network environments.

\subsection{Adversarial examples generation}
\label{AdvExpGen}
To craft adversarial samples, we use four different evasion attacks against each model. For this, the set of adversarial attacks that we use in the experiments is Fast Gradient Sign Method (FGSM), Projected Gradient Descent (PGD), Carlini \& Wagner (CW), and DeepFool (DF).

\begin{figure}[h!]
	\begin{center}
		\includegraphics[width=1\linewidth]{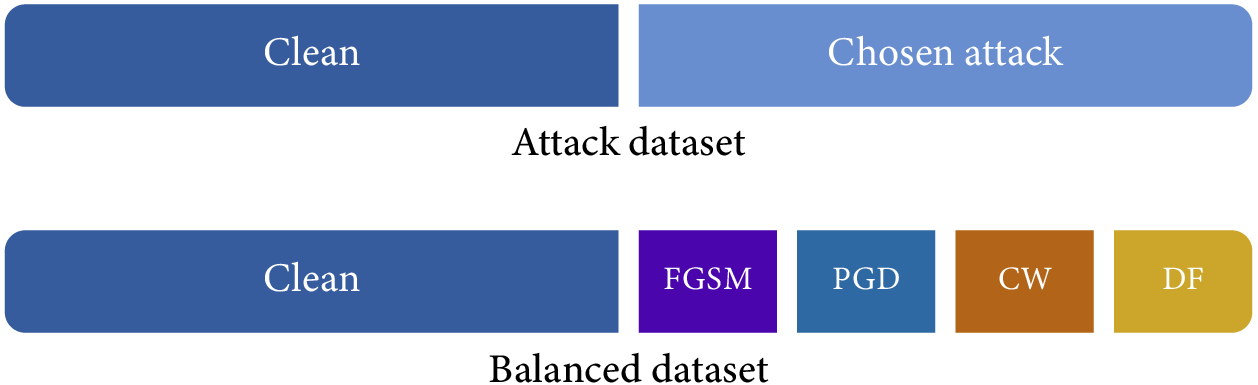}
		\caption{Balanced and attack-specific datasets composition}
		\label{Fig:datasets}
	\end{center}
\end{figure}

To train and evaluate the adversarial detectors, we create multiple training and testing datasets based on the network traffic dataset as shown in Fig. \ref{Fig:datasets}. We create an attack-specific dataset for each of the adversarial attacks, comprising the original dataset re-labeled as clean and of the same dataset altered by the chosen attack and labeled as adversarial. This results in balanced datasets between non-altered and altered samples. The final dataset that is created is a balanced version of the attacks. Indeed, while the balanced dataset still contained the original dataset, the altered part is equally shared between the four adversarial attacks.

It bears mentioning that our attacks are applied directly to feature vectors (feature-space attacks) as opposed to raw traffic (problem-space attacks) \cite{DBLP:conf/sp/PierazziPCC20}. Therefore, adversarial perturbations may not resemble truly realistic adversarial traffic \cite{DBLP:journals/corr/abs-2106-09380, DBLP:journals/jsac/HanWZCYLSY21}. However, in an attempt to preserve the functionality of our adversarial samples, certain semantic and syntactic constraints are taken into consideration \cite{merzouk2022investigating}. A set of features are kept unmodifiable, such as protocol type, service, or flags. We also limit the maximum perturbation rate to 10\% to ensure that the generated adversarial samples are close to the original samples.

\subsection{Serial DNN-IDS design}
The first step in this work is to design the DNN-IDS-serial. Its objective will be to classify benign and malicious traffic. Several design choices have to be tested before settling on an optimal design of the DNN: the number of hidden layers, the number of neurons per hidden layer, and the activation function.

The first design choice that was tested was the number of hidden layers to use. Indeed, while it was known that several hidden layers would be used in this DNN, the exact number of hidden layers is a tested parameter. We decided to test between two and four hidden layers. To test each configuration, we trained ten models of each configuration for fifty epochs each. During the training phase, we evaluated each model on our test dataset every ten epochs. This allowed us to keep only the best-performing model and avoid overfitting. We then averaged the performance obtained by each configuration. We found that using three ReLu-activated hidden layers was sufficient to achieve state-of-the-art performance on this dataset and that adding additional layers did not improve performance further. 

Based on these results, we decided to use three ReLu-activated hidden layers. We also varied the number of neurons in each hidden layer between forty and two hundred and fifty-six. This variation in the number of neurons resulted in almost no change in overall performance. It was therefore decided that each hidden layer would have two hundred and fifty-six neurons.

\subsection{Adversarial detectors design for DNN-IDS-serial }
\label{sec:compADS}
The choices made in this stage regarding our adversarial detectors revolved around the design of the detectors and their architecture. Regardless of these choices, we decided to give each detector the information of all layers placed before it in the DNN-IDS-serial. This means that $AdD_{n \in [1, N] }$ will be composed by the $[1, N]$ layers of DNN-IDS-serial in addition to the chosen detector design.

The design of the detectors refers to the number of layers as well as the number of neurons on each layer. We decided to test either one or two ReLu-activated layers, and either forty or two hundred and fifty-six neurons, as we did for our DNN-IDS-serial design.
The architecture itself refers to the transfer learning technique used to train our adversarial detectors. The tested architectures are the following:
\begin{itemize}
	\item Features Extraction: For $AdD_{n \in [1, N] }$, layers $[1, N]$ of the DNN-IDS-serial are shared with the adversarial detector. Those layers are made untrainable by freezing the weights and biases. To that set of layers are added our detector's layers. Those new layers are then trained to detect the attacks. Theoretically, this method allows faster training of the detectors. Indeed, since the shared layers are already trained on a similar task, the provided weights and biases should help to find optimal values for the detector's parameters faster than starting with random values.
	\item Fine-Tuning: For $AdD_{n \in [1, N] }$, layers $[1, N]$ of the DNN-IDS-serial are shared with the adversarial detector. To that set of layers are added our detector's layers. The ensemble is then trained to detect the adversarial attacks. Since this method changes the DNN-IDS-serial's weights and biases, there is a non-zero risk of changing the DNN-IDS-serial's performance. Since both tasks are quite similar, this change might be beneficial for the original detection performance. But it is also possible that we observe a drop in IDS performance after training our adversarial detectors.
	\item Duplication + Fine-Tuning: For $AdD_{n \in [1, N] }$, layers $[1, N]$ of the DNN-IDS-serial are duplicated and their copy is passed to the adversarial detector. To those layers are added our detector's layers. We then train the whole set of layers to detect adversarial attacks. This method benefits from the advantages of a Fine-Tuning architecture without risking a performance change for the DNN-IDS-serial. Indeed, since the layers are duplicated and not shared, the weights and biases of the original DNN-IDS-serial remain unchanged. This architecture would be a suitable solution if the performance change observed while using the Fine-Tuning architecture is detrimental to our DNN-IDS-serial.
\end{itemize}

To test the design and architecture of our adversarial detectors, we trained them for thirty epochs each and evaluated them in the same way as we evaluated the initial DNN-IDS-serial. The whole experiment was repeated 3 times to average out the results. FGSM attack was used to craft adversarial samples from the NSL-KDD dataset.
\begin{table}[th!]
	\centering
	\includegraphics[width=1\linewidth]{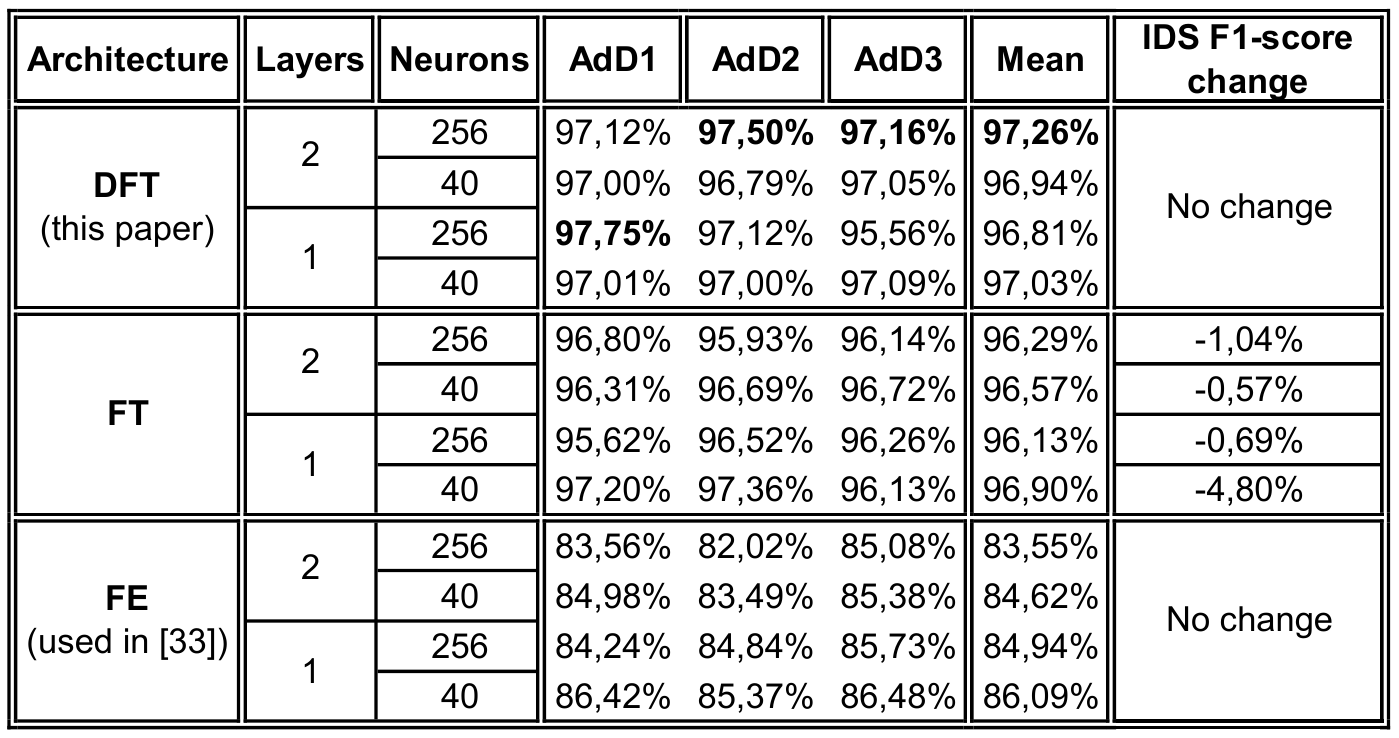}
	\caption{Comparaison of the detection rate of the adversarial detectors (AdD) in different design choices. For DFT and FE architectures, the IDS is not re-trained, hence no change in its performance.} 
	\label{tab:ftvsfe}
\end{table}

As shown in Table \ref{tab:ftvsfe},  we can see that the architectures allowing the re-training of the initial layers performed better at detecting the evasion attacks. This result is attributed to the fact that both tasks (detecting intrusion and evasion attacks) are quite similar. This means that the architecture allowing to fine-tune the DNN-IDS-serial benefits from its training and can capitalize on it. Unfortunately, an accuracy drop was observed in the DNN-IDS-serial when using the Fine-Tuning architecture. We notice also that the design choices of the adversarial detectors are of little influence on the detection rate. This is not surprising as it was already observed with the design of DNN-IDS-serial. We decided to use the Duplication + Fine-Tuning architecture with a design of 2 ReLu-activated layers of two-hundred and fifty-six neurons each for our adversarial detectors.

\subsection{Parallel DNN-IDS design}

\begin{figure}[th!]
	\begin{center}
		\includegraphics[width=\linewidth]{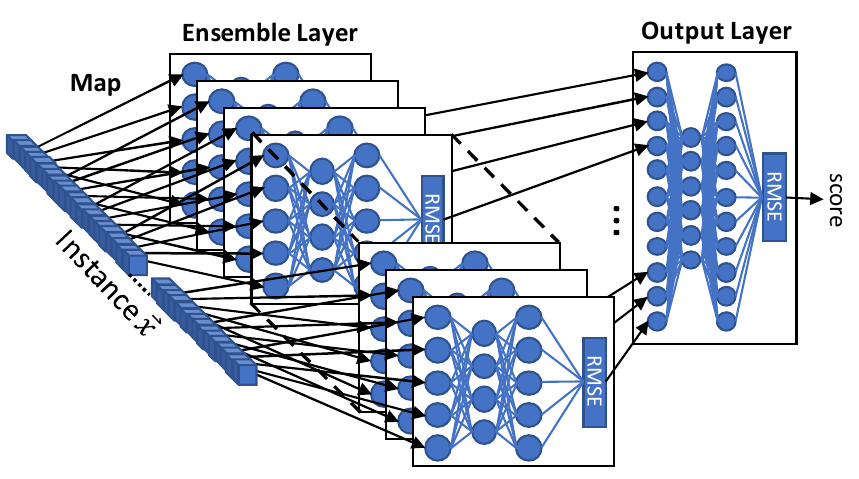}
		\caption{An illustration of Kitsune’s anomaly detection algorithm, from \cite{DBLP:conf/ndss/MirskyDES18}}
		\label{Fig:kistune}
	\end{center}
\end{figure}

In their original paper, the authors present Kitsune as a new IDS generation using an ensemble idea to split the learning process between multiple sub-models each training faster and using less computational and memory resources during the training process. To split the training process, the feature vector extracted from the training samples are first clustered together using their correlation to each other. Once clusters are decided, each cluster of features will be passed to one of the sub-models comprising the ensemble layer of Kitsune. The ensemble layer is comprised of a certain number of auto-encoders each taking one cluster of features and outputting the root-mean-square error (RMSE) between the original feature cluster and the one returned by the auto-encoder. Those RMSE are then used as input to another auto-encoder, the output layer, to compute the final RMSE which is then compared to a fixed threshold to perform a classification between benign and malicious samples. The Kitsune feature mapping, as well as ensemble \& output layers are shown in Fig. \ref{Fig:kistune}. The reason we chose this architecture is to add a deep inspection layer to better detect adversarial perturbations by increasing the granularity.

%In order to ensure that any differences in the results were brought by the change in the DNN-IDS design and not by a change in the experimentation process, we kept all the previous settings (attack parameters, fusion rules, datasets pre-processing). 

\subsubsection{Feature clustering}
As in the original Kitsune implementation, the network traffic features were split into multiple clusters. The number of clusters $K$ obtained during this step is bound by an arbitrarily fixed parameter (K is fixed to nine (09) for NSL-KDD and five (05) for CIC-IDS2017). Since the features present in our datasets are not the same as the ones used in the original Kitsune paper, the relationship between them is not as direct, and using the original Kitsune clustering method yielded unusable feature clusters. We decided to propose two clustering methods.

The first method, referred to as the distribution method, first performs the same clustering as the original Kitsune version. With highly uncorrelated features, the original Kitsune feature map may contain clusters containing only one feature. As passing only one feature to any neural network from the ensemble layer is not intended, the features belonging to those clusters are distributed between the other smaller clusters. This method will always yield clusters comprising a minimum of two features as well as a number of clusters inferior or equal to $K$.

The second method, referred to as the cut method, will first compute the number of features in each cluster by dividing the total amount of features by the number of clusters given by $K$. The features will then be ordered with regard to their correlation before being distributed between the clusters. This method will always yield $K$ clusters of approximately the same size (plus or minus one feature).

By training the DNN-IDS-parallel with the two clustering methods, we observed that the distribution method performed better than the cut method. Therefore, the distribution method was chosen as the reference method throughout the rest of the work.

In the original Kitsune implementation, the ensemble and output layers were composed of auto-encoders. we decided to replace those with deep neural networks similar to the DNN-IDS-serial used in the first part of this work. Each model in our ensemble and output layers is composed of three hidden ReLu layers of two hundred and fifty-six neurons and a bi-neuronal SoftMax layer. Each model of the ensemble layer was then trained for thirty epochs with a performance evaluation every ten epochs, keeping the best-performing model as the final trained model. From the predictions of those trained models for each instance, a new dataset was created. This new dataset was then used to train the model composing the output layer in the same manner as the models of the ensemble layer.
Using this configuration, the final model achieved state-of-the-art performance on our tested datasets.

\begin{figure}[t!]
	\begin{center}
		\includegraphics[width=1\linewidth]{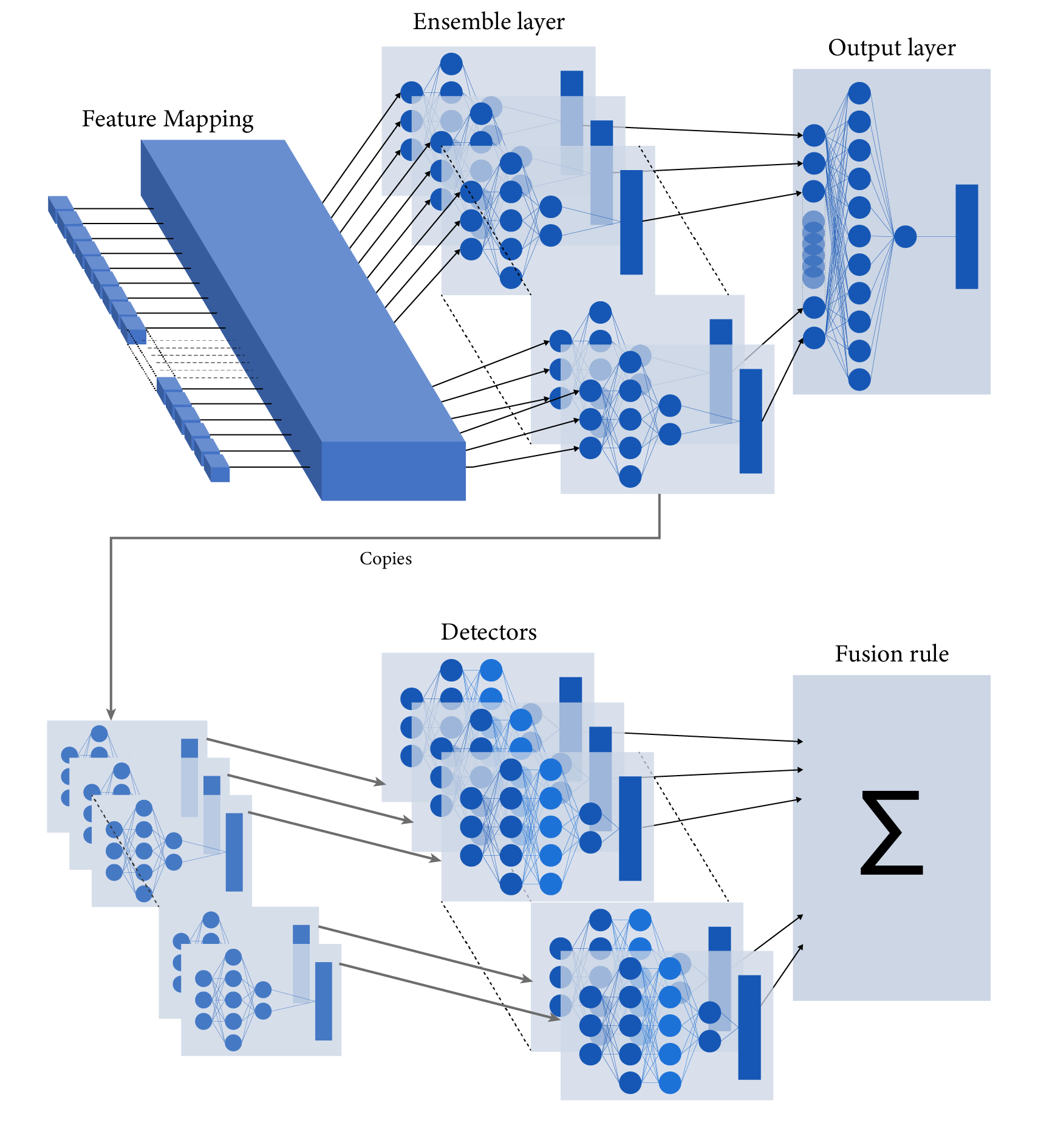}
		\caption{An illustration of the adversarial detectors in a parallel DNN-IDS design}
		\label{Fig:minato}
	\end{center}
\end{figure}

\subsection{Adversarial detectors design for DNN-IDS-parallel}

For the adversarial detector design, we decided to follow the same choices as for the DNN-IDS-serial. In Section \ref{sec:compADS}, we compared three architectures: Fine-tuning, Features extraction, and Duplication + Fine-tuning. The obtained results showed that being able to re-train any part of the DNN-IDS re-used or shared by the adversarial detectors led to a better detection rate. Since we also observed that using a Fine-Tuning architecture without duplication also led to a drop in accuracy for the DNN-IDS, we decided to use the Duplication + Fine-Tuning architecture. The multi adversarial detectors architecture in a parallel DNN-IDS design is illustrated in Fig. \ref{Fig:minato}. Each adversarial detector is composed of:

\begin{itemize}
	\item A copy of the input layer of the corresponding ensemble layer model.
	\item A copy of the hidden ReLu layers of the corresponding ensemble layer model, as well as their weights and biases.
	\item Two additional two hundred and fifty-six neurons hidden ReLu layers, randomly initialized.
	\item A mono-neuronal sigmoid output layer
\end{itemize}

The obtained array of detectors will then be trained in the same way as the ensemble layers: thirty epochs with an evaluation every ten epochs and keeping the best-evaluated detector.

\section{Performance evaluation}
\label{sec:fus}
This section presents the performance evaluation of the experimental settings. We first investigate the effect of the four adversarial attacks on the performance of the two DNN-IDSs. The adversarial detectors are then assessed by a cross-detection test. The final step is to evaluate the effectiveness of the three fusion rules. The results are then compared to another adversarial defense technique (adversarial training).

\subsection{Serial DNN-IDS}
\subsubsection{Adversarial attacks effect on DNN-IDS-serial}
In order to evaluate the impact of the four adversarial attacks, namely FGSM, PGD, CW, and DF on the performance of DNN-IDS-serial, we test the model with the four attack-specific datasets crafted as mentioned in Section \ref{AdvExpGen}.

\begin{figure}[h]
	%\centering
	\includegraphics[width=1\linewidth]{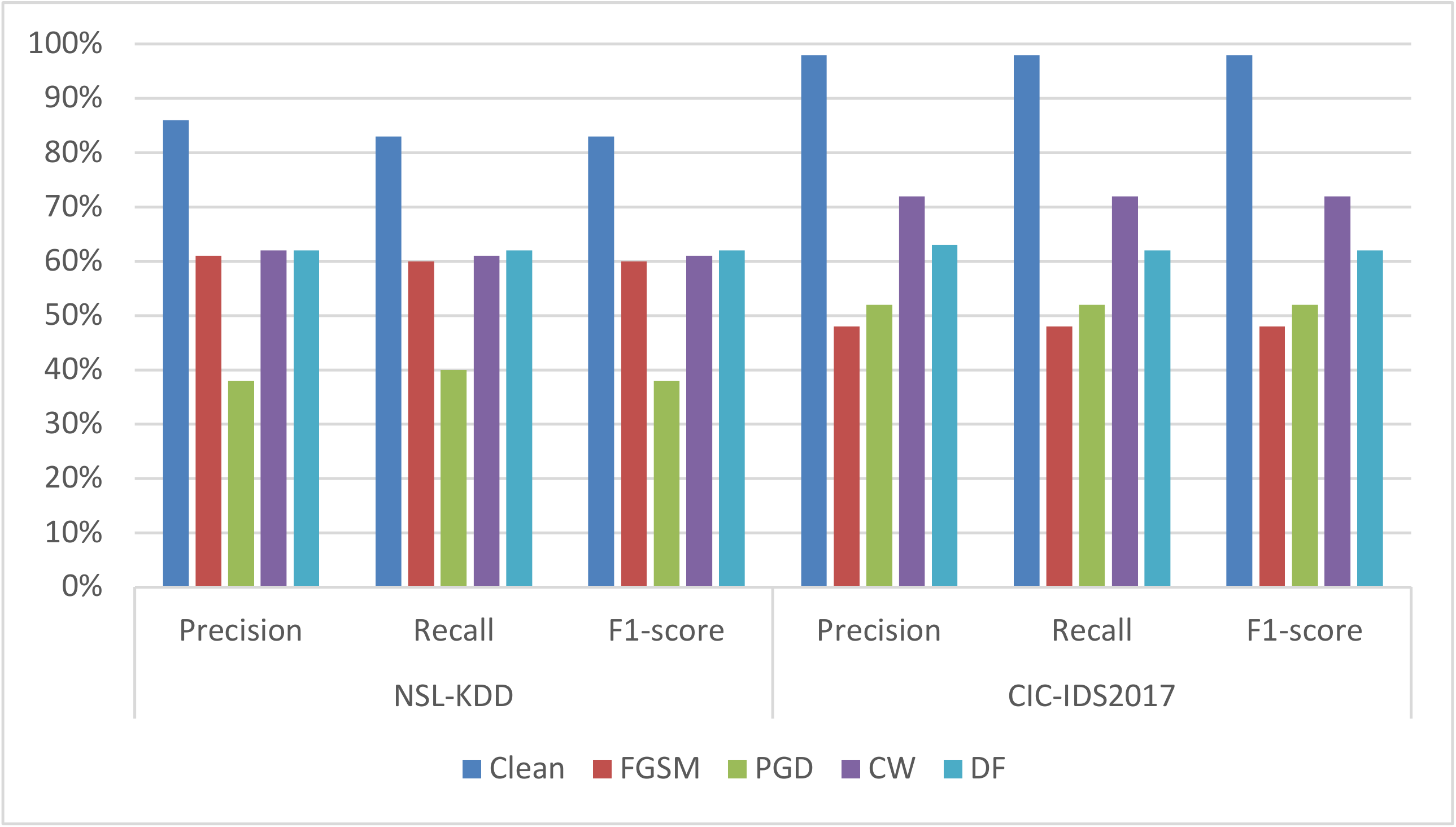}
	\caption{IDS performance difference between clean and adversarial network traffic in a serial DNN-IDS design}
	\label{fig:accDrop1}
\end{figure}

The results presented in Fig. \ref{fig:accDrop1} confirm the effect of adversarial attacks on DNN-IDS-serial. Indeed, we notice a significant drop in the IDS performance once the samples are altered by any of the adversarial attacks. For example, using PGD on the NSL-KDD dataset resulted in a decrease in F1-score from 83\% to 38\%, while using FGSM on the CIC-IDS2017 dataset resulted in a decrease in F1-score from 98\% to 48\%. These results are compatible with those found in the literature \cite{merzouk2022investigating, DBLP:conf/isncc/KhamisM20}. 

\subsubsection{Adversarial detectors performance}
To evaluate the performance of adversarial detectors, we conduct a cross-detection test. For this test, we train a set of adversarial detectors on one of our adversarial training datasets and then evaluate it against all adversarial testing datasets. This allows us to see how an adversarial detector trained on a specific attack or a specific subset of attacks would perform against an unknown attack. We also compare the results between detectors to analyze any differences in performance among them.

\begin{table*}[th!]
	\begin{center}
		\includegraphics[width=1\linewidth]{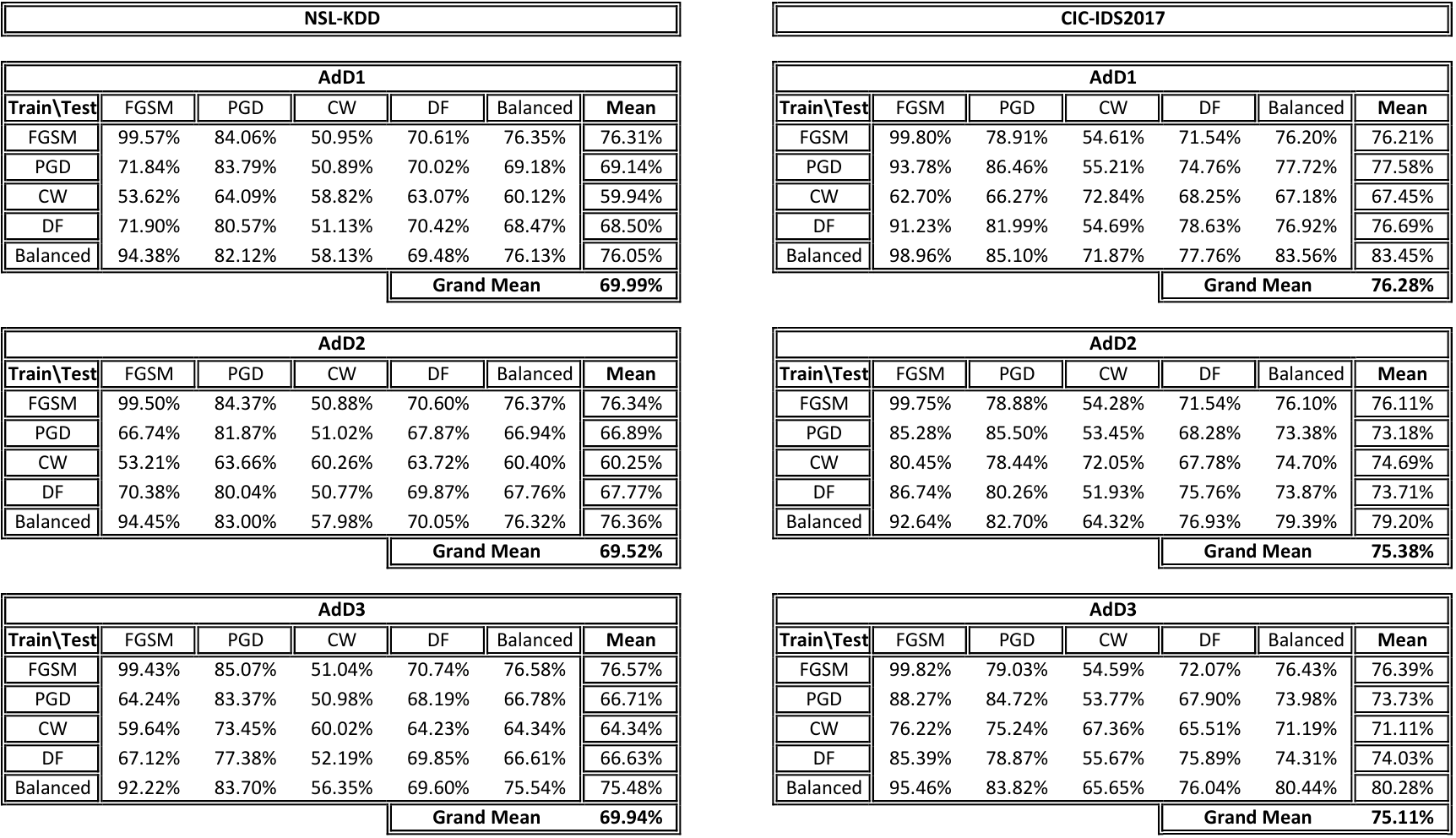}
		\caption{Detection rate of the individual adversarial detectors in a serial DNN-IDS design on NSL-KDD and CIC-IDS2017}
		\label{table: CrossIndv1}
	\end{center}
\end{table*}

As shown in Table \ref{table: CrossIndv1}, the first notable observation that can be made is that the detection rate differences between each detector in a set are minimal. If these minimal differences reflect the fact that every detector gives the same answer as the others for a vast majority of the samples, the fusion rules will yield results similar to those of any single detector. On the other hand, if those differences are only the results of the overall proportion of correctly identified samples, the fusion rules should yield better results than any single adversarial detector.

The detection rate results themselves reflect the fact that the CW attack is harder to detect than the other attacks. The detection rate difference could be amplified by the fact that our attack set comprises two FGSM-based attacks, FGSM itself and PGD, and that both could be detected similarly, resulting in a virtually unbalanced attack set. This hypothesis is also supported by the fact that the DF attack yields worst detection rates than both FGSM-based attacks. 

%It is also interesting to note that when facing a balanced set of attacks, while training on a balanced set of attacks is the best overall solution, training on FGSM attacked samples seems to yield close-to-best results. This, once again, could come from the fact that FGSM and PGD are based on similar principles. Implying that training on FGSM-altered samples yields a good detection rate for PGD-altered samples

\subsubsection{Fusion rules}

The final step is to implement the three fusion rules we decided to use: Majority Voting, Bayes Simple Average, and Dempster-Shafer Combination rules. To test these fusion rules, we use the same method as when testing the cross-detection of individual adversarial detectors. The only difference is that the decisions of each adversarial detector are now combined using the chosen rule to obtain a final decision.

\begin{table*}[h]
	\centering
	\includegraphics[width=1\linewidth]{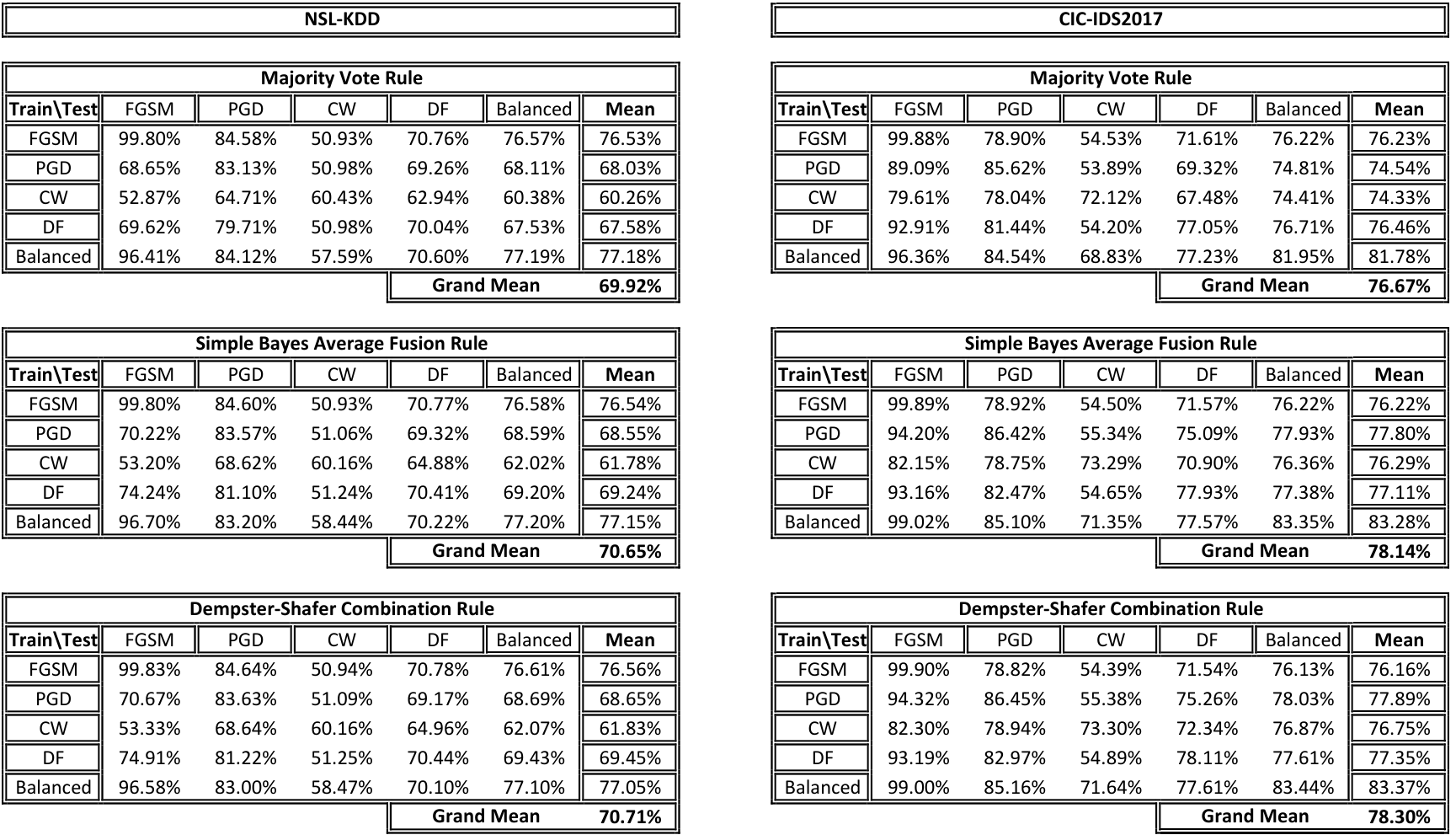}
	\caption{Detection rate of the fusion of the adversarial detectors in a serial DNN-IDS design on NSL-KDD and CIC-IDS2017}
	\label{tbl:tablefusionone1}
\end{table*}

As shown in Table \ref{tbl:tablefusionone1}, the three fusion rules produced similar results to those of the individual detectors. This confirms our previous hypothesis that each detector is probably giving the same response as the other detectors for most samples. Even though each detector has access to the information of an additional layer compared to its predecessor, it seems that all detectors are obtaining similar levels of information about the tested instances and therefore giving the same results. This means that the use of multiple adversarial detectors in a serial DNN-IDS design does not represent a significant improvement over the use of a single detector in our experimental setting.

\subsection{Parallel DNN-IDS}
\subsubsection{Adversarial attacks effect on DNN-IDS-parallel}

To perform each adversarial attack on our DNN-IDS-parallel, we decided to craft adversarial input for each model in the ensemble layer. To do this, each attack was performed against each specific subset of features used as input by that model.
\begin{figure}[th!]
	\centering
	\includegraphics[width=1\linewidth]{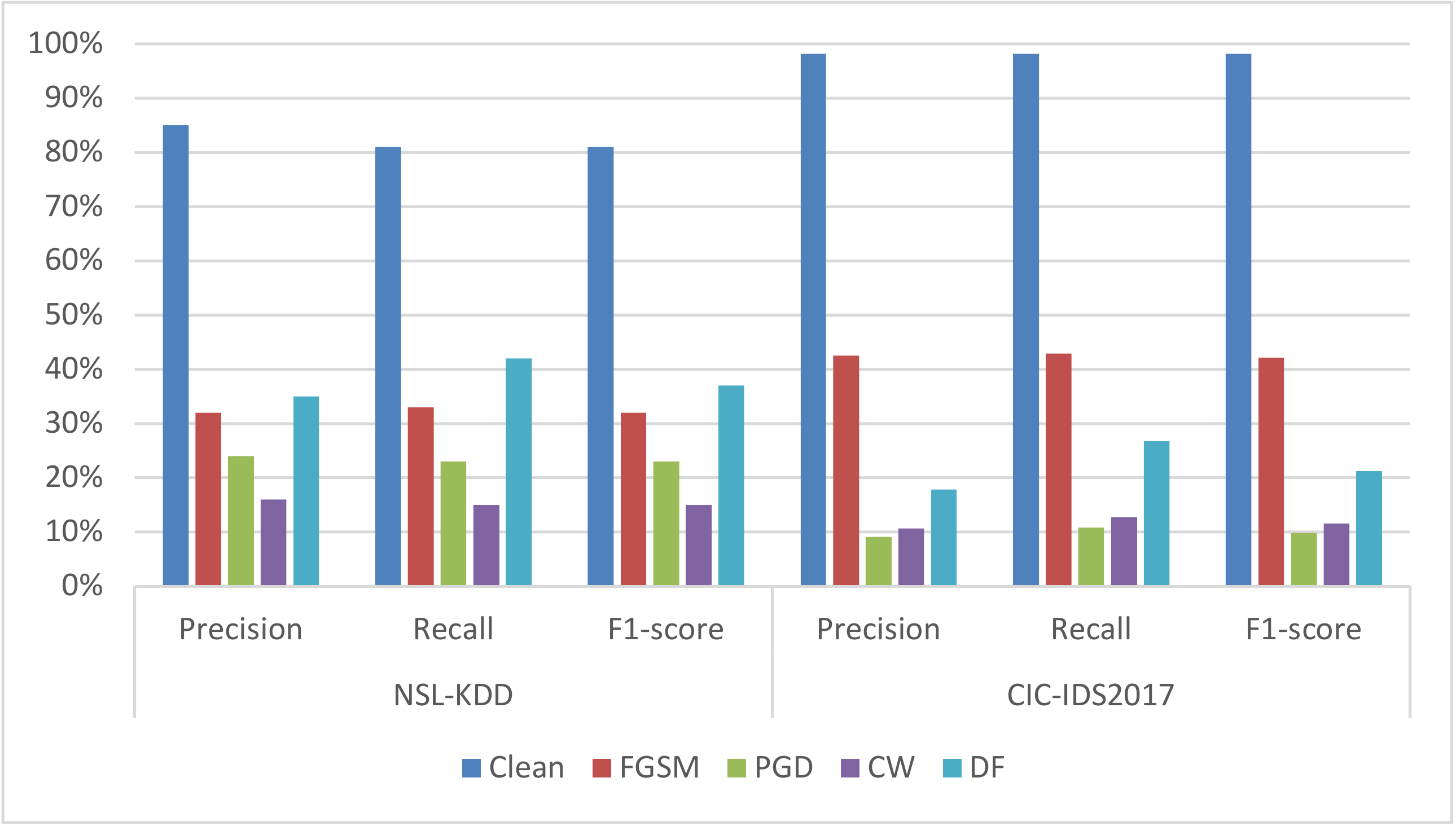}
	\caption{IDS accuracy difference between clean and adversarial network traffic in a parallel DNN-IDS design}
	\label{fig:perfDrop2}
\end{figure}

As shown in Fig. \ref{fig:perfDrop2}, similarly to the DNN-IDS-serial, we observe a significant drop in the IDS performance after performing adversarial attacks against our DNN-IDS-parallel. For example, using CW on the NSL-KDD dataset resulted in a decrease in F1-score from 81\% to 15\%, while using PGD on the CIC-IDS2017 dataset resulted in a decrease in F1-score from 98\% to 10\%. These results are compatible with those found in the literature \cite{DBLP:conf/ssci/ClementsYSHL21}. 

\subsubsection{Adversarial detectors performance}

When training the detectors on the various datasets, we observed greatly varying detection rates between the detectors as shown in Table \ref{tbl:table3}. Indeed, some detectors achieve an accuracy as low as 50\% while others can often reach as high as 100\% detection rate. We notice also that many individual detectors are having low detection rate against CW and DF attacks compared to FGSM and PGD. The differences in detection rate observed between the detectors are believed to come from the fact that most attacks alter specific features and disregard others. This would lead to some detectors rarely seeing any altered features, thus classifying almost every sample as legitimate and achieving only a 50\% detection rate. On the other hand, some detectors would receive the often-altered features and thus be able to detect an adversarial attack. As the best-performing detector will vary depending on the training dataset used, we expect that the use of multiple detectors will be an improvement over the use of a single detector.

\subsubsection{Fusion rules}
Considering the results obtained on individual detectors, it is now obvious that the way in which we will combine those individual classifications will be of great importance for the overall detection rate of our system. Indeed, from the three proposed fusion rules (cf. \ref{FR}), the majority vote rule is expected to be the least effective. Since some attacks only alter a small number of features, it is to be expected that only a few adversarial detectors will be able to detect these attacks while the majority will only see unaltered legitimate features, thus resulting in a false classification. While the Dempster-Shafer combination rule also uses the multiplicity of belief as a decisive factor, we believe that the effect will be less important than for the majority vote rule. The Dempster-Shafer combination rule only eliminates outcomes if they are absent from the possible outcomes decided by one of the detectors. This means that for this effect to eliminate one of the outcomes, it is required that one of the detectors returns a classification with a 100\% confidence. As the Dempster-Shafer combination rule is a more sophisticated rule and takes the detector uncertainty into consideration, it is expected that this rule will outperform both other rules.

The same cross-detection rate as before was computed, but this time combining the different classifications with each tested fusion rule. This method allows us to compare the different fusion rules but also to observe the possible generalization of detection through multiple attacks.

\begin{table*}[ht!]
	\centering
	\includegraphics[width=1\linewidth]{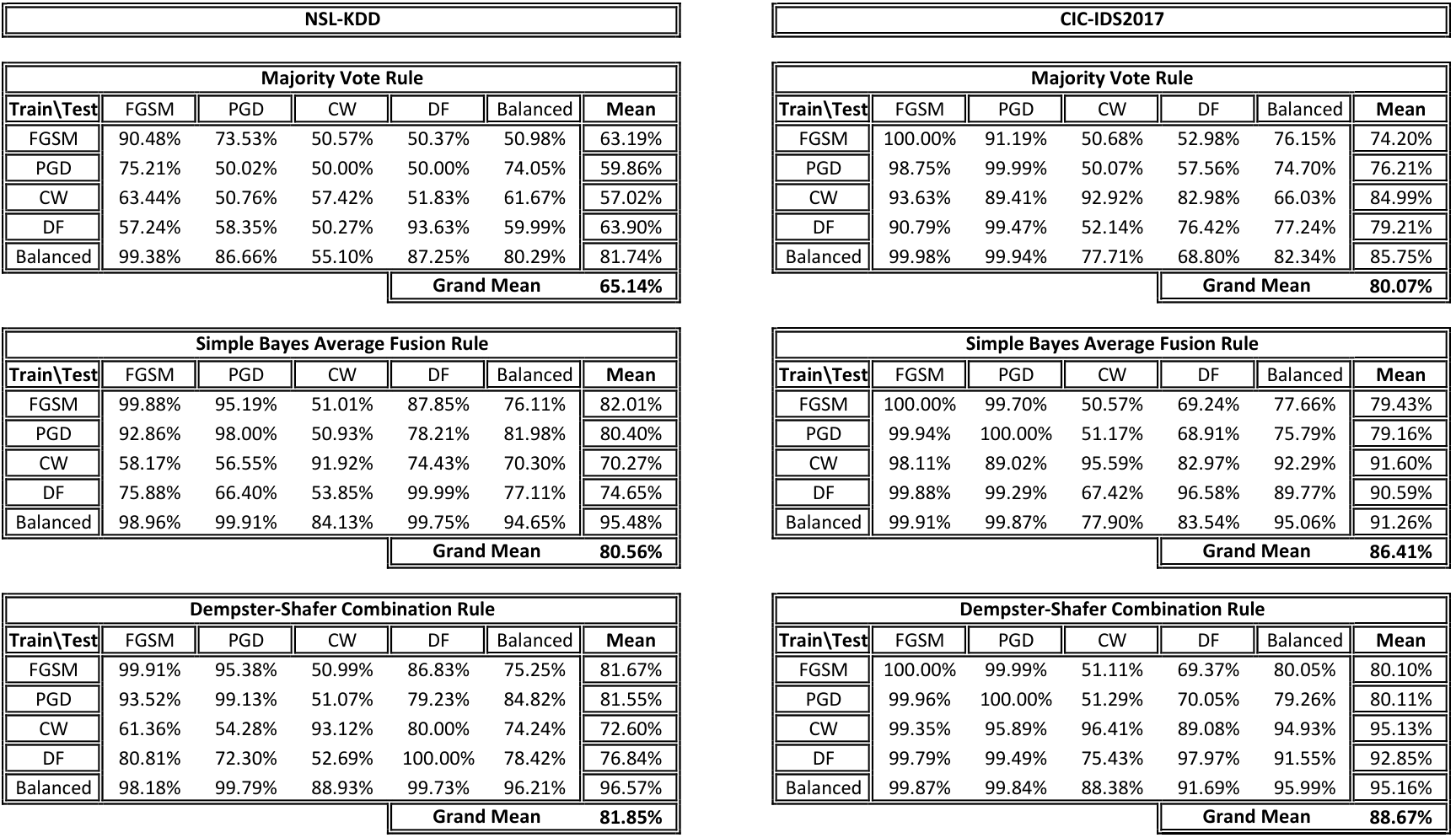}
	\caption{Detection rate of the fusion of the adversarial detectors in a parallel DNN-IDS design on NSL-KDD and CIC-IDS2017}
	\label{tbl:tablefusiontwo}
\end{table*}

\begin{table*}[hb!]
	\centering
	\includegraphics[width=1\linewidth]{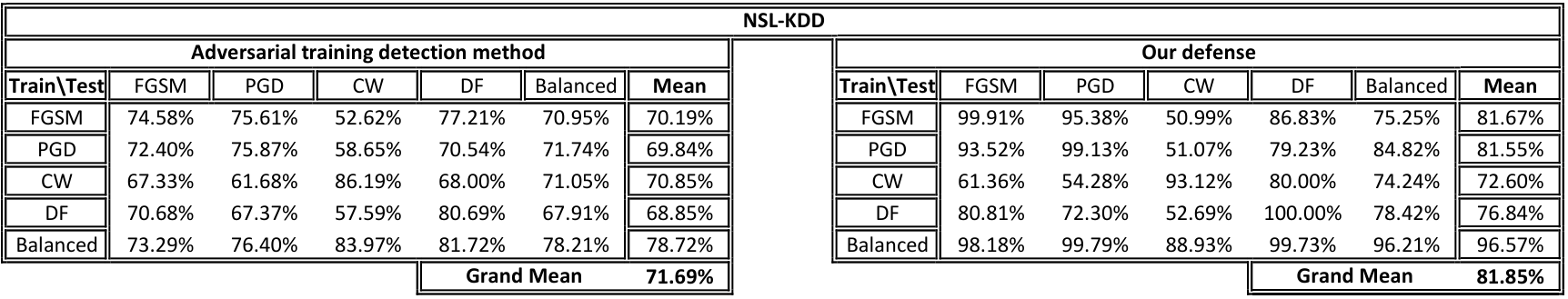}
	\caption{Detection rate comparison between adversarial training and our defense in a parallel DNN-IDS design on NSL-KDD dataset}
	\label{tbl:CompAdvTrNSL}
\end{table*}
\begin{table*}[hb!]
	\centering
	\includegraphics[width=1\linewidth]{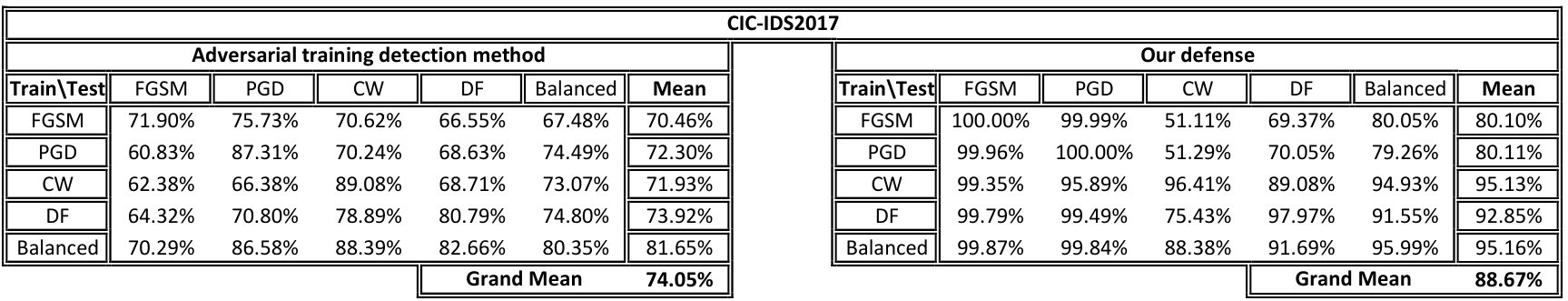}
	\caption{Detection rate comparison between adversarial training and our defense in a parallel DNN-IDS design on CIC-IDS2017 dataset}
	\label{tbl:CompAdvTrCIC}
\end{table*}

\begin{table*}[h!]
	\centering
	\includegraphics[height=0.95\textheight]{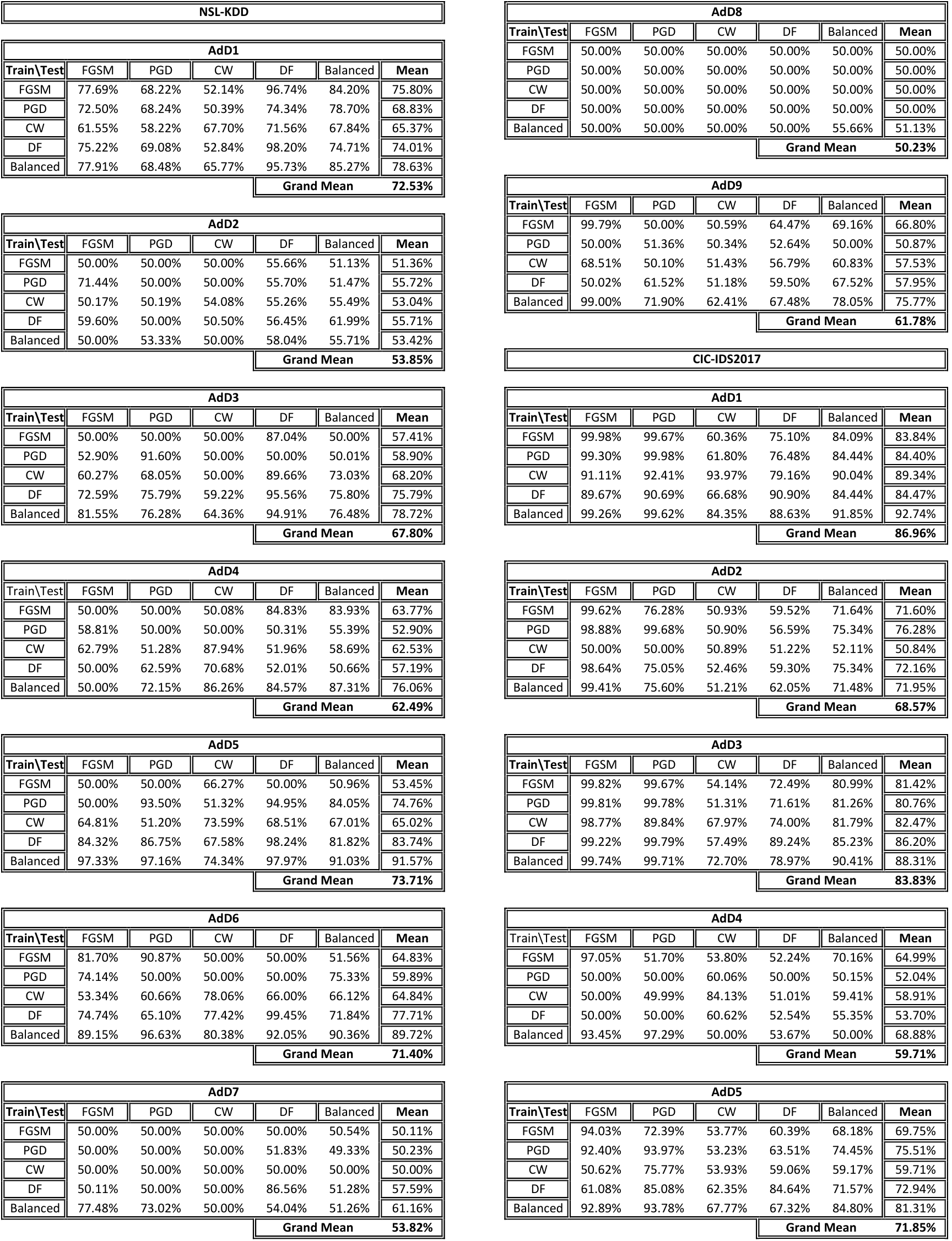}
	\caption{Detection rate of the individual adversarial detectors in a parallel DNN-IDS design on NSL-KDD and CIC-IDS2017}
	\label{tbl:table3}
\end{table*}

From Table \ref{tbl:tablefusiontwo}, we notice that the fusion of the adversarial detectors using the Dempster-Shafer combination rule is outperforming the individual detectors in terms of detection rate. This confirms the relevance of using multiple, strategically placed adversarial detectors. 

The contrast in observed results could come from the differences in the way every attack is performed. Indeed, both the CW and DF attacks approach the feature alteration in a way that is different from the two other, more similar, attacks (i.e., FGSM and PGD). The fact that using the Dempster-Shafer fusion rule or the Simple Bayes Average fusion rule leads to overall better results than the Majority Vote fusion rule confirms the importance of taking the contextual information into account.

\subsection{Comparison with existing defensive strategies}

Our proposed defense falls into the category of anomaly detection, as explained in Section \ref{sec:AnomalyDetection}. There are other detection methods, some of which use robust classifiers as detectors. Indeed, it is possible to detect adversarial samples by comparing the classification of two models: a baseline model, trained only on non-adversarial samples, and a robust adversarial model trained on both non-adversarial and adversarial samples. When subjecting a sample to the two models, one can assume that if the two models classify it differently, the sample is adversarial. In theory, if the submitted sample is non-adversarial, the base model and the robust model should classify it correctly. On the other hand, if the sample is adversarial, the base model should classify it incorrectly, while the robust model should classify it correctly.

We implemented this defense by training a second parallel DNN-IDS model on a dataset consisting of equal amounts of adversarial and non-adversarial samples. We then submitted a likewise test dataset to both models and recorded their predictions. By comparing these two sets of predictions, we obtained a final set of predictions classifying each sample as adversarial or non-adversarial. As shown in Table \ref{tbl:CompAdvTrNSL} and Table \ref{tbl:CompAdvTrCIC}, this method yielded a detection rate of 71.69\% and 74.05\% for NSL-KDD and CIC-IDS2017 respectively. As we demonstrated earlier, our proposal using the Dempster-Shafer combination rule obtained a strictly higher detection rate on both datasets.

These differences in results can be explained by the contrasting objectives of the two strategies. Our proposed defense is specifically aimed at detecting adversarial samples. On the other hand, the adversarial training detection method is a by-product of training a robust classifier which should remove the need for detection (in theory, at least). In addition, the difference in result could stem from the fact that our proposed defense has a higher level of granularity by inspecting each subset of features separately and then reaching a consensus decision using an appropriate fusion rule.

\section{Conclusion and Future Work}
\label{sec:cfw}

In this paper, we proposed a new defense approach for evasion attacks against network-based intrusion detection systems. To evaluate it, we implemented two deep learning-based IDS models (one serial and one parallel) and assessed the effect of four known adversarial attacks on their performance, namely: Fast Gradient Sign Method, Projected Gradient Descent, Carlini \& Wagner, and DeepFool. A transfer learning technique was employed in the design of the adversarial detection scheme to enable a better information flow between the IDS and the adversarial detectors. Simulation of zero-day attack scenarios allowed for a more reliable evaluation of the performance of the proposed defense when exposed to evasion attacks that were not seen in the training phase. To further improve the detection rate, we proposed and investigated an alternative to a single detector by using multiple strategically placed detectors combined with a sophisticated fusion rule. We show that combining multiple detectors can further improve the detection rate over a single detector in a parallel IDS design. The reported results confirm the relevance of the proposed defense with respect to existing techniques in the literature.

In future work, we would like to improve the fusion rules used in our design. Indeed, while the use of the Dempster-Shafer fusion rule gave some promising results, it was not fully successful in all cases, which could mean that a more sophisticated fusion rule is needed to model the complex relationship between detectors classification. In addition, we would like to further investigate the relationship between detectability and the strength of adversarial attacks.

\section*{Declaration of competing interest}
The authors declare that they have no known competing financial interests or personal relationships that could have appeared to influence the work reported in this paper.

\section*{CRediT authorship contribution statement}
\textbf{Islam Debicha:} Conceptualization, Validation, Methodology, Writing - Original Draft. \textbf{Richard Bauwens:} Software, Visualization, Writing - Original Draft. \textbf{Thibault Debatty:} Validation, Writing - Review \& Editing, Supervision. \textbf{Jean-Michel Dricot:} Validation, Writing - Review \& Editing, Supervision. \textbf{Tayeb Kenaza:} Validation, Writing - Review \& Editing, Supervision. \textbf{Wim Mees:} Funding acquisition, Supervision.

%\section*{References}

\bibliography{bibliography.bib}

%\newpage

\bigskip 

\piccaptioninside

\parpic[l]{\includegraphics[width=0.3\linewidth, height=0.2\textwidth]{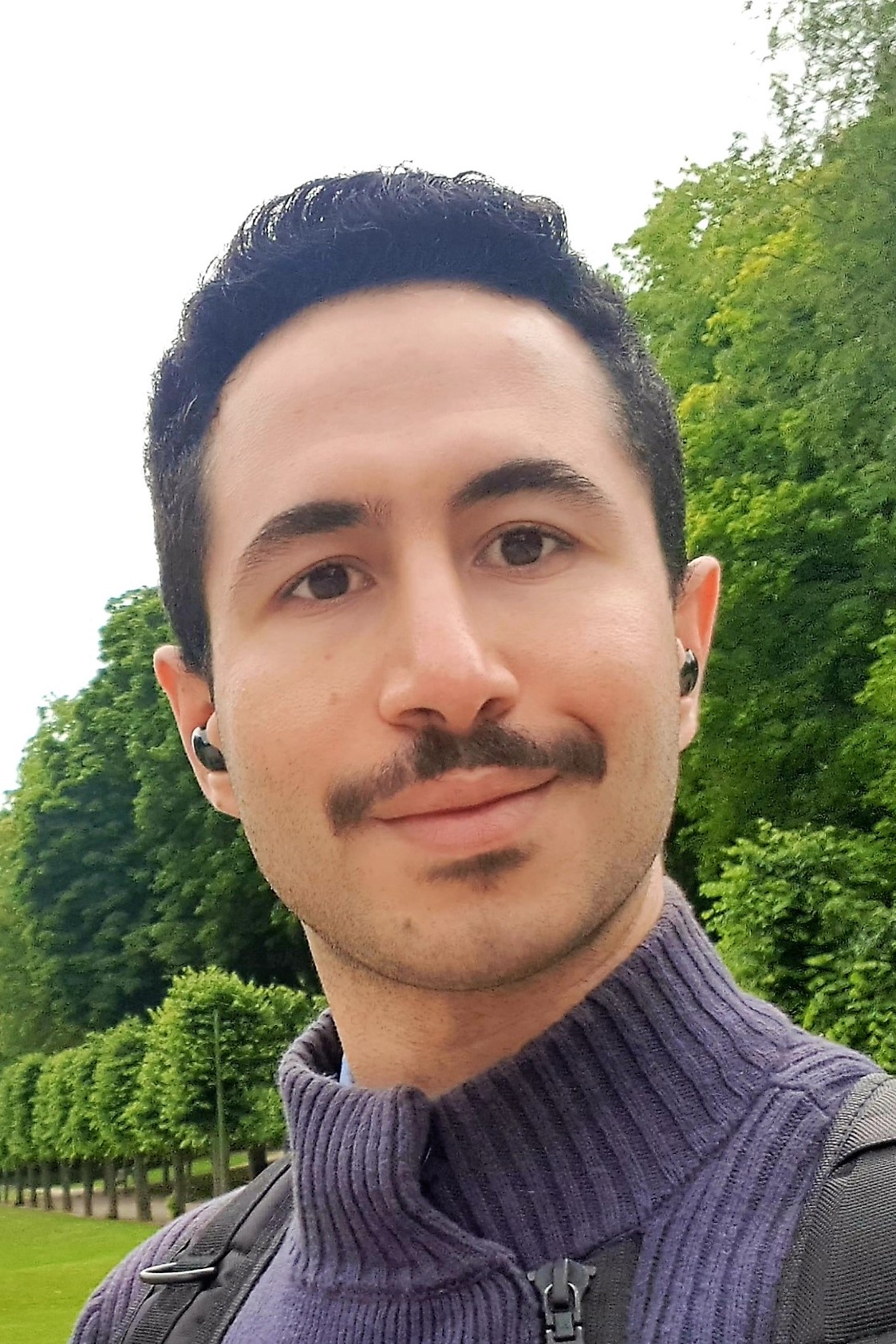}}

%------------------------------------------
\textbf{Islam Debicha} received his master's degree in Computer Science with a focus in network security in 2018. He is pursuing a joint Ph.D. in machine learning-based intrusion detection systems at ULB and ERM, Brussels, Belgium. He works at ULB cybersecurity Research Center and Cyber Defence Lab on evasion attacks against machine learning-based intrusion detection systems. He is the author or co-author of 5 peer-reviewed scientific publications. His research interests include defenses against adversarial examples, machine learning applications in cybersecurity, data fusion, and network security.

\piccaptioninside

\parpic[l]{\includegraphics[width=0.3\linewidth, height=0.2\textwidth]{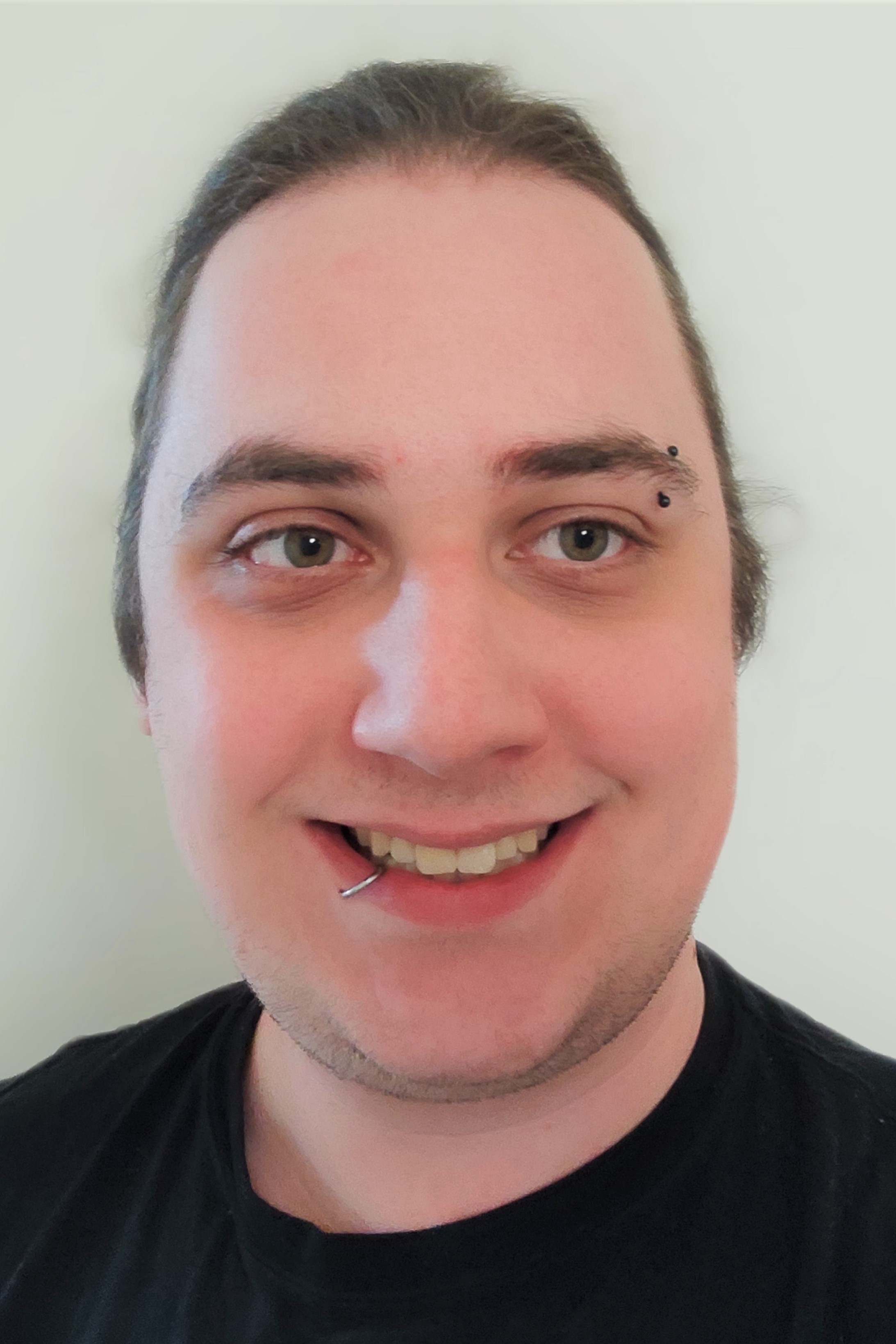}}
\textbf{Richard Bauwens} received his Master's degree in Cybersecurity with a focus on System Design and Analysis from the Université Libre de Bruxelles (ULB). He works on security automation and threat detection for a private company in Brussels. His work and research interests include security automation, fileless malware, automatic threat detection and handling, and adversarial learning.

\piccaptioninside

\parpic[l]{\includegraphics[width=0.3\linewidth, height=0.2\textwidth]{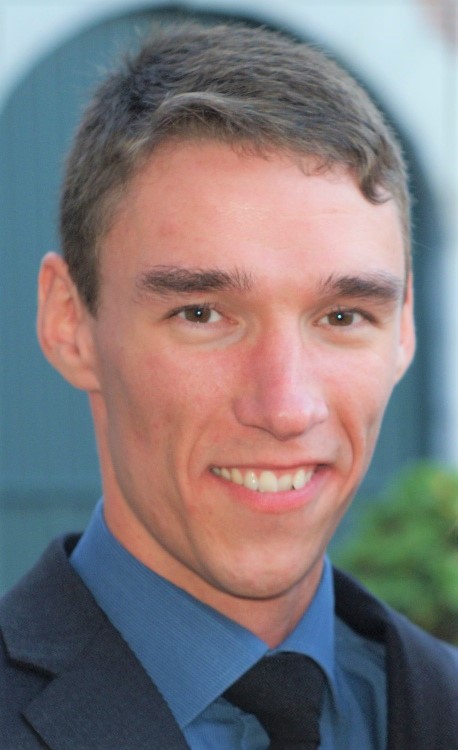}}
\textbf{Thibault Debatty} obtained a master's degree in applied engineering sciences at the Royal Military Academy (RMA) in Belgium, followed by a master's degree in applied computer science at the Vrije Universiteit Brussel (VUB). Finally, he obtained a Ph.D. with a specialization in distributed computing at both Telecom Paris and the RMA. He is now an associate professor at the RMA, where he teaches courses in networking, distributed information systems, and information security. He is also president of the jury of the Master of Science in Cybersecurity organized by the Université Libre de Bruxelles (ULB), the RMA, and four other institutions.

\piccaptioninside

\parpic[l]{\includegraphics[width=0.3\linewidth, height=0.2\textwidth]{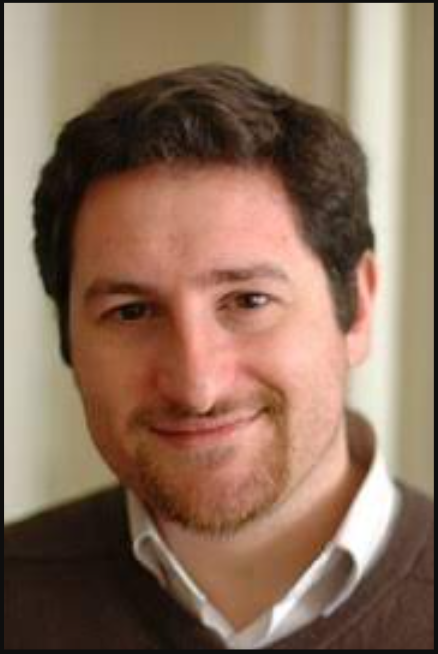}}
\textbf{Jean-Michel Dricot} leads research on network security with a specific focus on the IoT and wireless networks. He teaches communication networks, mobile networks, the internet of things, and network security. Prior to his tenure at the ULB, Jean-Michel Dricot obtained a Ph.D. in network engineering, with a focus on wireless sensor network protocols and architectures. In 2010, Jean-Michel Dricot was appointed professor at the Université Libre de Bruxelles, with tenure in mobile and wireless networks. He is the author or co-author of more than 100+ papers published in peer-reviewed international Journals and Conferences and served as a reviewer for European projects.

\piccaptioninside

\parpic[l]{\includegraphics[width=0.3\linewidth, height=0.2\textwidth]{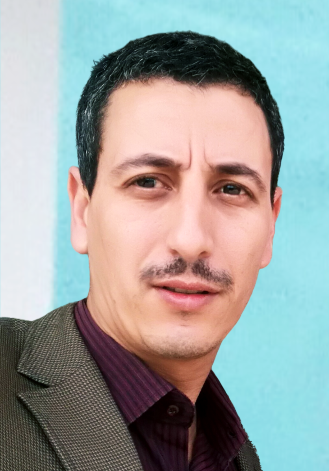}}
\textbf{Tayeb Kenaza }received a Ph.D. degree in computer science from Artois University, France, in 2011. Since 2017, he is the Head of the Computer Security Laboratory at the Military Polytechnic School of Algiers. He is currently a senior lecturer in Computer Science Department at the same school. His research and publication interests include Computer Network Security, Wireless Communication Security, Intelligent Systems, and Data Mining.

\piccaptioninside

\parpic[l]{\includegraphics[width=0.3\linewidth, height=0.2\textwidth]{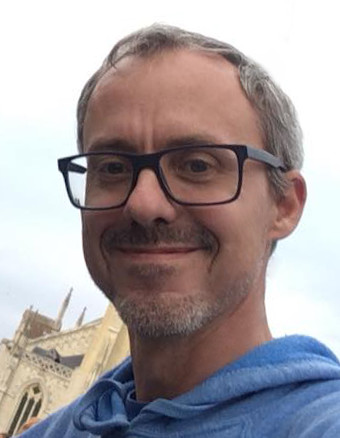}}
\textbf{Wim Mees} is Professor in computer science and cyber security at the Royal Military Academy and is leading the Cyber Defence Lab. He is also teaching in the Belgian inter-university Master in Cybersecurity, and in the Master in Enterprise Architecture at the IC Institute. Wim has participated in and coordinated numerous national and European research projects as well EDA and NATO projects and task groups. 
\end{document}